\def\sphinxdocclass{article}
\documentclass[letterpaper,12pt,english]{sphinxhowto}
\ifdefined\pdfpxdimen
   \let\sphinxpxdimen\pdfpxdimen\else\newdimen\sphinxpxdimen
\fi \sphinxpxdimen=.75bp\relax

\PassOptionsToPackage{warn}{textcomp}
\usepackage[utf8]{inputenc}
\ifdefined\DeclareUnicodeCharacter
 \ifdefined\DeclareUnicodeCharacterAsOptional
  \DeclareUnicodeCharacter{"00A0}{\nobreakspace}
  \DeclareUnicodeCharacter{"2500}{\sphinxunichar{2500}}
  \DeclareUnicodeCharacter{"2502}{\sphinxunichar{2502}}
  \DeclareUnicodeCharacter{"2514}{\sphinxunichar{2514}}
  \DeclareUnicodeCharacter{"251C}{\sphinxunichar{251C}}
  \DeclareUnicodeCharacter{"2572}{\textbackslash}
 \else
  \DeclareUnicodeCharacter{00A0}{\nobreakspace}
  \DeclareUnicodeCharacter{2500}{\sphinxunichar{2500}}
  \DeclareUnicodeCharacter{2502}{\sphinxunichar{2502}}
  \DeclareUnicodeCharacter{2514}{\sphinxunichar{2514}}
  \DeclareUnicodeCharacter{251C}{\sphinxunichar{251C}}
  \DeclareUnicodeCharacter{2572}{\textbackslash}
 \fi
\fi
\usepackage{cmap}
\usepackage[T1]{fontenc}
\usepackage{amsmath,amssymb,amstext}
\usepackage{babel}
\usepackage{times}
\usepackage[Bjarne]{fncychap}
\usepackage{sphinx}
\sphinxsetup{verbatimwithframe=false, VerbatimColor={rgb}{0.95,0.95,0.95}}
\usepackage{geometry}

\usepackage{hyperref}
\usepackage{hypcap}
\addto\captionsenglish{}
\addto\captionsenglish{}
\addto\captionsenglish{\renewcommand{\literalblockname}{Listing}}

\addto\captionsenglish{\renewcommand{\literalblockcontinuedname}{continued from previous page}}
\addto\captionsenglish{\renewcommand{\literalblockcontinuesname}{continues on next page}}

\addto\extrasenglish{}

\usepackage{txfontsb}

\title{EPIC Documentation}
\date{Sep 17, 2018}
\release{1.4}
\author{Rafael J. F. Marcondes}
\newcommand{\sphinxlogo}{\sphinxincludegraphics{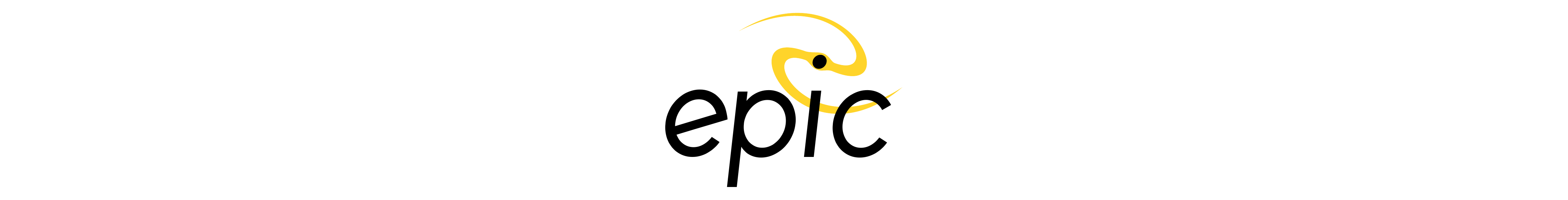}\par}
\renewcommand{\releasename}{Release}
\makeindex

\begin{document}

\maketitle
\sphinxtableofcontents
\phantomsection\label{\detokenize{index::doc}}

\section{Welcome to the EPIC user’s guide!}
\label{\detokenize{welcome:welcome-to-the-epic-user-s-guide}}\label{\detokenize{welcome::doc}}
Easy Parameter Inference in Cosmology
(EPIC) is my implementation in Python of a MCMC code for
Bayesian inference of parameters of cosmological models and model comparison via
the computation of Bayesian evidences.

\subsection{Why EPIC?}
\label{\detokenize{welcome:why-epic}}
I started to develop EPIC as a means of learning how parameter inference can be
made with Markov Chain Monte Carlo, rather than trying to decipher other
codes or using them as black boxes.
The program has fulfilled this purposed and went on to incorporate a few
cosmological observables that I have actually employed in some of my
publications. Now I release this code in the hope it can be useful to
students seeking to learn some of the methods used in Observational Cosmology
and even to use it for their own work.
It still lacks some important features.
A Boltzmann solver is not available.
It is possible that I will integrate it with CLASS %
\begin{footnote}[1]\sphinxAtStartFootnote
Lesgourgues, J. “The Cosmic Linear Anisotropy Solving System (CLASS) I: Overview”. arXiv:1104.2932 {[}astro-ph.IM{]}; Blas, D., Lesgourgues, J., Tram, T. “The Cosmic Linear Anisotropy Solving System (CLASS). Part II: Approximation schemes”. Journal of Cosmology and Astroparticle Physics 07 (2011) 034.
\end{footnote} to make it more
useful for advanced research.
Stay tuned for more.

On the other hand, development is active and with recent versions it is now
possible not only to use EPIC’s Cosmology Calculator but also run MCMC
simulations from a nice graphical interface.
You can also, check out these other features:
\begin{itemize}
\item {} 
Cross-platform: the code runs on Python 3 in any operating system.

\item {} 
EPIC features a Cosmology Calculator that also supports a few models other
than the standard \(\Lambda\text{CDM}\) model.
The list of models include interacting dark energy and some dark energy
equation-of-state parametrizations.
The code can output some key distance calculations and compare them between
different models over a range of redshifts, generating extensively
customizable plots.

\item {} 
It uses Python’s \sphinxcode{\sphinxupquote{multiprocessing}} library for evolution of chains in
parallel in MCMC simulations.
The separate processes can communicate with each other through some
\sphinxcode{\sphinxupquote{multiprocessing}} utilities, which made possible the implementation of the
Parallel Tempering algorithm. %
\begin{footnote}[2]\sphinxAtStartFootnote
Removed in current version. If you need to use Parallel Tempering, please use version 1.0.4 of EPIC.
\end{footnote} This method is capable of detecting
and accurately sampling posterior distributions that present two or more
separated peaks.

\item {} 
Convergence between independent chains is tested with the multivariate
version of the Gelman and Rubin test, a very robust method.

\item {} 
Also, the plots are beautiful and can be customized to a great extent
directly from the command line or from the graphical interface, without
having to change the code.
You can view triangle plots with marginalized distributions of parameters,
predefined derived parameters, two-dimensional joint-posterior distributions,
autocorrelation plots, cross-correlation plots, sequence plots, convergence
diagnosis and more.

\end{itemize}

Try it now!

\section{How to install}
\label{\detokenize{howtoinstall:how-to-install}}\label{\detokenize{howtoinstall::doc}}
You can obtain this program either from PyPi (installing with \sphinxcode{\sphinxupquote{pip}}) or cloning the public repository from BitBucket.
But first, it is recommended that you make these changes inside a
virtual environment.

\subsection{Setting up a virtual environment}
\label{\detokenize{howtoinstall:setting-up-a-virtual-environment}}

\subsubsection{On Unix systems}
\label{\detokenize{howtoinstall:on-unix-systems}}
The preferred way is using Python3’s \sphinxcode{\sphinxupquote{venv}} module, available in Python 3.3
and superior versions.
When choosing this option on Unix systems, the main script \sphinxcode{\sphinxupquote{epic.py}} will be
executable from any location.
In a directory of your preference, create a virtual Python environment (for
example, named \sphinxcode{\sphinxupquote{EPIC-env}}) and activate it with:

\fvset{hllines={, ,}}%
\begin{sphinxVerbatim}[commandchars=\\\{\}]
\PYGZdl{} python3 \PYGZhy{}m venv EPIC\PYGZhy{}env
\PYGZdl{} source EPIC\PYGZhy{}env/bin/activate
\end{sphinxVerbatim}

When you finish using the environment and want to leave it you can just use
\sphinxcode{\sphinxupquote{\$ deactivate}}.
To activate it again, which you need in a new session, just run the activation
command above (the second line only).
More details about Python3’s \sphinxcode{\sphinxupquote{venv}}
\sphinxhref{https://docs.python.org/3/library/venv.html}{here}.

Alternatively, you can install
\sphinxhref{https://github.com/pyenv}{pyenv and pyenv-virtualenv}, which let you create
a virtual environment and even choose another version of Python to install.
This is done with:

\fvset{hllines={, ,}}%
\begin{sphinxVerbatim}[commandchars=\\\{\}]
\PYGZdl{} pyenv virtualenv 3.6.1 EPIC\PYGZhy{}env \PYGZsh{} or choose other version you like.
\PYGZdl{} pyenv activate EPIC\PYGZhy{}env \PYGZsh{} use \PYGZsq{}pyenv deactivate\PYGZsq{} to deactivate it.
\end{sphinxVerbatim}

Note that this version of EPIC is not compatible with Python 2 anymore.

\subsubsection{On Windows}
\label{\detokenize{howtoinstall:on-windows}}
EPIC is supported on Windows through the Conda system.
Download and install \sphinxhref{https://conda.io/miniconda.html}{Miniconda3}.
Then, from the Anaconda prompt, create and activate your environment with:

\fvset{hllines={, ,}}%
\begin{sphinxVerbatim}[commandchars=\\\{\}]
\PYGZdl{} conda create \PYGZhy{}n EPIC\PYGZhy{}env python
\PYGZdl{} conda activate EPIC\PYGZhy{}env
\end{sphinxVerbatim}

\subsection{Defining your custom EPIC home folder (optional)}
\label{\detokenize{howtoinstall:defining-your-custom-epic-home-folder-optional}}
By default, data from the MCMC simulations will be saved to a folder named
\sphinxcode{\sphinxupquote{simulations}} in the same location of the input \sphinxcode{\sphinxupquote{.ini}} file used in each
simulation.
You can have a common location for all results by defining an environment
variable.
This makes sense since the output files can be big and you might want to store
them in a separate driver.

To do this, before installing EPIC, set the variable \sphinxcode{\sphinxupquote{EPIC\_USER\_PATH}} to the
location of your preference.
On Unix systems, add the following line to your \sphinxcode{\sphinxupquote{\textasciitilde{}/.bashrc}} or
\sphinxcode{\sphinxupquote{\textasciitilde{}/.bashprofile}} or \sphinxcode{\sphinxupquote{\textasciitilde{}/.profile}} file:

\fvset{hllines={, ,}}%
\begin{sphinxVerbatim}[commandchars=\\\{\}]
\PYG{n}{export} \PYG{n}{EPIC\PYGZus{}USER\PYGZus{}PATH}\PYG{o}{=}\PYG{l+s+s2}{\PYGZdq{}}\PYG{l+s+s2}{/path/to/folder/}\PYG{l+s+s2}{\PYGZdq{}}
\end{sphinxVerbatim}

so the variable will be defined in every session.
On Windows, open Control Panel, go to System and Security, System, Advanced
system settings, Environment Variables.
In the section User variables for (your user), click the New button, browse to
the directory that you want to set as EPIC’s home and create the variable with
the name \sphinxcode{\sphinxupquote{EPIC\_USER\_PATH}}.

\subsection{Installing EPIC}
\label{\detokenize{howtoinstall:installing-epic}}

\subsubsection{From PyPi}
\label{\detokenize{howtoinstall:from-pypi}}
The easiest way to install this program is to get it from PyPi.
It is also the most recommended since it can be easily updated when new
versions come out.
Inside your virtual environment, run:

\fvset{hllines={, ,}}%
\begin{sphinxVerbatim}[commandchars=\\\{\}]
\PYGZdl{} pip install epic\PYGZhy{}code
\end{sphinxVerbatim}

You are now good to go.
During the installation, some \sphinxcode{\sphinxupquote{.ini}} files will be extracted to the EPIC’s home folder in your home folder or in \sphinxcode{\sphinxupquote{EPIC\_USER\_PATH}} if you defined it before.
If you used \sphinxcode{\sphinxupquote{venv}} and your system is Unix, you may be able to launch EPIC’s
graphical interface just by running:

\fvset{hllines={, ,}}%
\begin{sphinxVerbatim}[commandchars=\\\{\}]
\PYGZdl{} epic.py
\end{sphinxVerbatim}

from any location. In other configurations, when this cannot be achieved, the
script \sphinxcode{\sphinxupquote{epic.py}} will be exported to the home folder so you can execute it
from that location with:

\fvset{hllines={, ,}}%
\begin{sphinxVerbatim}[commandchars=\\\{\}]
\PYGZdl{} python epic.py
\end{sphinxVerbatim}

To check and install updates if available, just run:

\fvset{hllines={, ,}}%
\begin{sphinxVerbatim}[commandchars=\\\{\}]
\PYGZdl{} pip install \PYGZhy{}\PYGZhy{}upgrade epic\PYGZhy{}code
\end{sphinxVerbatim}

from inside your environment.

\subsubsection{Cloning the git repository}
\label{\detokenize{howtoinstall:cloning-the-git-repository}}
If you plan to contribute to this program you can clone the git repository at
\sphinxurl{https://bitbucket.org/rmarcondes/epic}, running:

\fvset{hllines={, ,}}%
\begin{sphinxVerbatim}[commandchars=\\\{\}]
\PYGZdl{} git clone https://bitbucket.org/rmarcondes/epic.git
\end{sphinxVerbatim}

After downloading the repository, \sphinxcode{\sphinxupquote{cd}} into the \sphinxcode{\sphinxupquote{epic}} folder and install
the program with:

\fvset{hllines={, ,}}%
\begin{sphinxVerbatim}[commandchars=\\\{\}]
\PYGZdl{} pip install \PYGZhy{}e .
\end{sphinxVerbatim}

This should be repeated as you edit the program unless you are always running
it from the \sphinxcode{\sphinxupquote{EPIC}} folder.
To use modified code from the python interactive interpreter you need to install
it again.

\section{The Cosmology calculator}
\label{\detokenize{cosmology:cosmology-module}}\label{\detokenize{cosmology::doc}}\label{\detokenize{cosmology:the-cosmology-calculator}}
Starting with version 1.1, there is now a module available that makes it easy
for the user to perform calculations on the background cosmology given a
specific model.
A few classes of models are predefined. Each of these have also defined what
are their components, but also allowing some variations.
For example, the \(\Lambda\text{CDM}\) model requires at least cold dark
matter (\sphinxcode{\sphinxupquote{cdm}}) and cosmological constant (\sphinxcode{\sphinxupquote{lambda}}), but one can also
include baryonic fluid (\sphinxcode{\sphinxupquote{baryons}}), or treat both matter components together
by specifying the composed species \sphinxcode{\sphinxupquote{matter}}.
It is also possible to include \sphinxcode{\sphinxupquote{photons}} or \sphinxcode{\sphinxupquote{radiation}}.
For models of interaction between dark matter and dark energy (\sphinxcode{\sphinxupquote{cde}}), the
former is labelled \sphinxcode{\sphinxupquote{idm}} and the latter \sphinxcode{\sphinxupquote{ide}} or \sphinxcode{\sphinxupquote{ilambda}} in the case
that its equation-of-state parameter is still \(-1\) (in which the model is
then labelled \sphinxcode{\sphinxupquote{cde\_lambda}}) rather than a free parameter or presents
evolution described by some function.
The models available and their components specification are defined in the file
\sphinxcode{\sphinxupquote{EPIC/cosmology/model\_recipes.ini}}. The fluids contained in this file are in
turn defined in \sphinxcode{\sphinxupquote{EPIC/cosmology/available\_species.ini}}, where properties like
the type of equation of state, the parameters and other relevant informations
are set.

In version 1.2, a graphical user interface (GUI) for this cosmology calculator
has been added to EPIC.
After describing how all calculations are done in EPIC, we present the GUI in
the last section.

\subsection{The interactive calculator}
\label{\detokenize{cosmology_module:the-interactive-calculator}}\label{\detokenize{cosmology_module::doc}}
We start (preferably on jupyter notebook) importing the module and
creating our cosmology object:

\fvset{hllines={, ,}}%
\begin{sphinxVerbatim}[commandchars=\\\{\}]
\PYG{g+gp}{\PYGZgt{}\PYGZgt{}\PYGZgt{} }\PYG{k+kn}{from} \PYG{n+nn}{EPIC}\PYG{n+nn}{.}\PYG{n+nn}{cosmology} \PYG{k}{import} \PYG{n}{cosmic\PYGZus{}objects} \PYG{k}{as} \PYG{n}{cosmo}
\PYG{g+gp}{\PYGZgt{}\PYGZgt{}\PYGZgt{} }\PYG{n}{LCDM} \PYG{o}{=} \PYG{n}{cosmo}\PYG{o}{.}\PYG{n}{CosmologicalSetup}\PYG{p}{(}\PYG{l+s+s1}{\PYGZsq{}}\PYG{l+s+s1}{lcdm}\PYG{l+s+s1}{\PYGZsq{}}\PYG{p}{)}
\end{sphinxVerbatim}

Only the label of the model is really needed here, since the essentials
are already predefined in the program, as mentioned above. With this,
one can explore the properties assigned to the object. For example,
\sphinxcode{\sphinxupquote{LCDM.model}} will print \sphinxcode{\sphinxupquote{lcdm}}. \sphinxcode{\sphinxupquote{LCDM.species}} is a dictionary of
\sphinxcode{\sphinxupquote{Fluid}} objects identified by the components labels, in this case
\sphinxcode{\sphinxupquote{cdm}} and \sphinxcode{\sphinxupquote{lambda}}. There is also a dedicated class for an
equation-of-state parameter or function, which becomes an attribute of
its fluid. We can assess its value, type, etc.
\sphinxcode{\sphinxupquote{LCDM.species{[}'lambda'{]}.EoS.value}} will print \sphinxcode{\sphinxupquote{-1}}.

But let us proceed in a slightly different way, setting up our model
with some options. Since we predominantly work with flat cosmologies (in
fact, curvature is not supported yet in the current version), the
flatness condition is imposed in the density parameter of one of the
fluids. We will choose the dark energy density parameter to be the
derived parameter, but we could have chosen dark matter as well. Also,
by default, the code prefers to work with physical densities (for
example \(\Omega_{c0} h^2\)) rather than the common
\(\Omega_{c0}\). You can change this with the option
\sphinxcode{\sphinxupquote{physical=False}}. We will add the radiation and matter fluids. Note
that this will override the optional inclusion of baryons and remove
them, if given. The radiation fluid is parametrized by the temperature
of the cosmic microwave background. The model will have three free
parameters: the physical density parameter of matter
(\(\Omega_{m0}h^2\)), the CMB temperature (\(T_{\gamma}\), which
we usually keep fixed) and the Hubble parameter \(h\); and one
derived parameter, which is the density parameter of the cosmological
constant, \(\Omega_{\Lambda}h^2\).

\fvset{hllines={, ,}}%
\begin{sphinxVerbatim}[commandchars=\\\{\}]
\PYG{g+gp}{\PYGZgt{}\PYGZgt{}\PYGZgt{} }\PYG{n}{LCDM} \PYG{o}{=} \PYG{n}{cosmo}\PYG{o}{.}\PYG{n}{CosmologicalSetup}\PYG{p}{(}
\PYG{g+gp}{... }   \PYG{l+s+s1}{\PYGZsq{}}\PYG{l+s+s1}{lcdm}\PYG{l+s+s1}{\PYGZsq{}}\PYG{p}{,}
\PYG{g+gp}{... }   \PYG{n}{optional\PYGZus{}species}\PYG{o}{=}\PYG{p}{[}\PYG{l+s+s1}{\PYGZsq{}}\PYG{l+s+s1}{baryons}\PYG{l+s+s1}{\PYGZsq{}}\PYG{p}{,} \PYG{l+s+s1}{\PYGZsq{}}\PYG{l+s+s1}{radiation}\PYG{l+s+s1}{\PYGZsq{}}\PYG{p}{]}\PYG{p}{,} \PYG{c+c1}{\PYGZsh{} baryons will be ignored because of \PYGZsq{}matter\PYGZsq{} in the line bellow}
\PYG{g+gp}{... }   \PYG{n}{combined\PYGZus{}species}\PYG{o}{=}\PYG{p}{[}\PYG{l+s+s1}{\PYGZsq{}}\PYG{l+s+s1}{matter}\PYG{l+s+s1}{\PYGZsq{}}\PYG{p}{]}\PYG{p}{,}
\PYG{g+gp}{... }   \PYG{n}{derived}\PYG{o}{=}\PYG{l+s+s1}{\PYGZsq{}}\PYG{l+s+s1}{lambda}\PYG{l+s+s1}{\PYGZsq{}}
\PYG{g+gp}{... }\PYG{p}{)}
\end{sphinxVerbatim}

We can then obtain the solution to the background cosmology with EPIC.

\subsubsection{Solving the background cosmology}
\label{\detokenize{cosmology_module:solving-the-background-cosmology}}
It is as simple as this:

\fvset{hllines={, ,}}%
\begin{sphinxVerbatim}[commandchars=\\\{\}]
\PYG{g+gp}{\PYGZgt{}\PYGZgt{}\PYGZgt{} }\PYG{n}{LCDM}\PYG{o}{.}\PYG{n}{solve\PYGZus{}background}\PYG{p}{(}\PYG{n}{accepts\PYGZus{}default}\PYG{o}{=}\PYG{k+kc}{True}\PYG{p}{)}
\end{sphinxVerbatim}

Normally, a set of parameters would be given to this function in the
form of a dictionary with the parameters’ labels as keys, like in
\sphinxcode{\sphinxupquote{parameter\_space=\{'Oc0': 0.26, 'Ob0': 0.048, 'Or0':8e-5, 'H0':67.8\}}}.
However, we can also ommit it and turn on the option \sphinxcode{\sphinxupquote{accepts\_default}}
and then the default values defined in the
\sphinxcode{\sphinxupquote{EPIC/cosmology/default\_parameter\_values.ini}} file will be used for
the parameters. Next, we plot the energy densities and density
parameters. Here I do it in a jupyter notebook with the help of this
simple function below:

\fvset{hllines={, ,}}%
\begin{sphinxVerbatim}[commandchars=\\\{\}]
\PYG{g+gp}{\PYGZgt{}\PYGZgt{}\PYGZgt{} }\PYG{o}{\PYGZpc{}}\PYG{n}{matplotlib} \PYG{n}{inline}
\PYG{g+gp}{\PYGZgt{}\PYGZgt{}\PYGZgt{} }\PYG{k+kn}{import} \PYG{n+nn}{matplotlib}\PYG{n+nn}{.}\PYG{n+nn}{pyplot} \PYG{k}{as} \PYG{n+nn}{plt}
\PYG{g+gp}{\PYGZgt{}\PYGZgt{}\PYGZgt{} }\PYG{k+kn}{import} \PYG{n+nn}{numpy} \PYG{k}{as} \PYG{n+nn}{np}
\PYG{g+gp}{\PYGZgt{}\PYGZgt{}\PYGZgt{} }\PYG{n}{plt}\PYG{o}{.}\PYG{n}{rcParams}\PYG{p}{[}\PYG{l+s+s1}{\PYGZsq{}}\PYG{l+s+s1}{text.usetex}\PYG{l+s+s1}{\PYGZsq{}}\PYG{p}{]} \PYG{o}{=} \PYG{k+kc}{True}
\PYG{g+gp}{\PYGZgt{}\PYGZgt{}\PYGZgt{} }\PYG{n}{plt}\PYG{o}{.}\PYG{n}{rcParams}\PYG{p}{[}\PYG{l+s+s1}{\PYGZsq{}}\PYG{l+s+s1}{font.size}\PYG{l+s+s1}{\PYGZsq{}}\PYG{p}{]} \PYG{o}{=} \PYG{l+m+mi}{16}
\PYG{g+gp}{\PYGZgt{}\PYGZgt{}\PYGZgt{} }\PYG{n}{plt}\PYG{o}{.}\PYG{n}{rcParams}\PYG{p}{[}\PYG{l+s+s1}{\PYGZsq{}}\PYG{l+s+s1}{figure.dpi}\PYG{l+s+s1}{\PYGZsq{}}\PYG{p}{]} \PYG{o}{=} \PYG{l+m+mi}{144}

\PYG{g+gp}{\PYGZgt{}\PYGZgt{}\PYGZgt{} }\PYG{k}{def} \PYG{n+nf}{show\PYGZus{}densities}\PYG{p}{(}\PYG{n}{model}\PYG{p}{,} \PYG{o}{*}\PYG{o}{*}\PYG{n}{kwargs}\PYG{p}{)}\PYG{p}{:}
\PYG{g+gp}{... }   \PYG{k+kn}{from} \PYG{n+nn}{EPIC}\PYG{n+nn}{.}\PYG{n+nn}{cosmology} \PYG{k}{import} \PYG{n}{rho\PYGZus{}critical\PYGZus{}SI}
\PYG{g+gp}{... }   \PYG{n}{hubble} \PYG{o}{=} \PYG{n}{model}\PYG{o}{.}\PYG{n}{HubbleParameter}\PYG{o}{.}\PYG{n}{get\PYGZus{}value}\PYG{p}{(}\PYG{o}{*}\PYG{o}{*}\PYG{n}{kwargs}\PYG{p}{)}
\PYG{g+gp}{... }   \PYG{n}{rho\PYGZus{}cr0} \PYG{o}{=} \PYG{n}{rho\PYGZus{}critical\PYGZus{}SI}\PYG{p}{(}
\PYG{g+gp}{... }       \PYG{n}{hubble} \PYG{o}{*} \PYG{p}{(}\PYG{l+m+mi}{100} \PYG{k}{if} \PYG{n}{model}\PYG{o}{.}\PYG{n}{physical\PYGZus{}density\PYGZus{}parameters} \PYG{k}{else} \PYG{l+m+mi}{1}\PYG{p}{)}
\PYG{g+gp}{... }   \PYG{p}{)}
\PYG{g+gp}{... }   \PYG{n}{fig}\PYG{p}{,} \PYG{n}{ax} \PYG{o}{=} \PYG{n}{plt}\PYG{o}{.}\PYG{n}{subplots}\PYG{p}{(}\PYG{l+m+mi}{1}\PYG{p}{,} \PYG{l+m+mi}{2}\PYG{p}{)}
\PYG{g+gp}{... }   \PYG{n}{fig}\PYG{o}{.}\PYG{n}{set\PYGZus{}size\PYGZus{}inches}\PYG{p}{(}\PYG{l+m+mi}{12}\PYG{p}{,} \PYG{l+m+mi}{4}\PYG{p}{)}
\PYG{g+gp}{... }   \PYG{k}{for} \PYG{n}{key} \PYG{o+ow}{in} \PYG{n}{model}\PYG{o}{.}\PYG{n}{background\PYGZus{}solution\PYGZus{}Omegas}\PYG{o}{.}\PYG{n}{keys}\PYG{p}{(}\PYG{p}{)}\PYG{p}{:}
\PYG{g+gp}{... }       \PYG{n}{ax}\PYG{p}{[}\PYG{l+m+mi}{0}\PYG{p}{]}\PYG{o}{.}\PYG{n}{plot}\PYG{p}{(}\PYG{n}{model}\PYG{o}{.}\PYG{n}{a\PYGZus{}range}\PYG{p}{,} \PYG{n}{model}\PYG{o}{.}\PYG{n}{background\PYGZus{}solution\PYGZus{}Omegas}\PYG{p}{[}\PYG{n}{key}\PYG{p}{]}\PYG{p}{,} \PYG{n}{lw}\PYG{o}{=}\PYG{l+m+mi}{2}\PYG{p}{,} \PYG{n}{label}\PYG{o}{=}\PYG{n}{key}\PYG{p}{)}
\PYG{g+gp}{... }       \PYG{n}{ax}\PYG{p}{[}\PYG{l+m+mi}{1}\PYG{p}{]}\PYG{o}{.}\PYG{n}{plot}\PYG{p}{(}\PYG{n}{model}\PYG{o}{.}\PYG{n}{a\PYGZus{}range}\PYG{p}{,} \PYG{n}{rho\PYGZus{}cr0} \PYGZbs{}
\PYG{g+gp}{... }                  \PYG{o}{*} \PYG{p}{(}\PYG{n}{hubble}\PYG{o}{*}\PYG{o}{*}\PYG{o}{\PYGZhy{}}\PYG{l+m+mi}{2} \PYG{k}{if} \PYG{n}{model}\PYG{o}{.}\PYG{n}{physical\PYGZus{}density\PYGZus{}parameters} \PYG{k}{else} \PYG{l+m+mi}{1}\PYG{p}{)} \PYGZbs{}
\PYG{g+gp}{... }                  \PYG{o}{*} \PYG{n}{model}\PYG{o}{.}\PYG{n}{background\PYGZus{}solution\PYGZus{}rhos}\PYG{p}{[}\PYG{n}{key}\PYG{p}{]}\PYG{p}{,} \PYG{n}{lw}\PYG{o}{=}\PYG{l+m+mi}{2}\PYG{p}{,} \PYG{n}{label}\PYG{o}{=}\PYG{n}{key}\PYG{p}{)}
\PYG{g+gp}{... }   \PYG{n}{ax}\PYG{p}{[}\PYG{l+m+mi}{0}\PYG{p}{]}\PYG{o}{.}\PYG{n}{set\PYGZus{}xscale}\PYG{p}{(}\PYG{l+s+s1}{\PYGZsq{}}\PYG{l+s+s1}{log}\PYG{l+s+s1}{\PYGZsq{}}\PYG{p}{)}
\PYG{g+gp}{... }   \PYG{n}{ax}\PYG{p}{[}\PYG{l+m+mi}{0}\PYG{p}{]}\PYG{o}{.}\PYG{n}{set\PYGZus{}ylabel}\PYG{p}{(}\PYG{l+s+sa}{r}\PYG{l+s+s1}{\PYGZsq{}}\PYG{l+s+s1}{\PYGZdl{}}\PYG{l+s+s1}{\PYGZbs{}}\PYG{l+s+s1}{Omega\PYGZdl{}}\PYG{l+s+s1}{\PYGZsq{}}\PYG{p}{)}
\PYG{g+gp}{... }   \PYG{n}{ax}\PYG{p}{[}\PYG{l+m+mi}{0}\PYG{p}{]}\PYG{o}{.}\PYG{n}{set\PYGZus{}xlabel}\PYG{p}{(}\PYG{l+s+sa}{r}\PYG{l+s+s1}{\PYGZsq{}}\PYG{l+s+s1}{\PYGZdl{}a\PYGZdl{}}\PYG{l+s+s1}{\PYGZsq{}}\PYG{p}{)}
\PYG{g+gp}{... }   \PYG{n}{ax}\PYG{p}{[}\PYG{l+m+mi}{0}\PYG{p}{]}\PYG{o}{.}\PYG{n}{grid}\PYG{p}{(}\PYG{n}{which}\PYG{o}{=}\PYG{l+s+s1}{\PYGZsq{}}\PYG{l+s+s1}{both}\PYG{l+s+s1}{\PYGZsq{}}\PYG{p}{,} \PYG{n}{linestyle}\PYG{o}{=}\PYG{l+s+s1}{\PYGZsq{}}\PYG{l+s+s1}{:}\PYG{l+s+s1}{\PYGZsq{}}\PYG{p}{)}
\PYG{g+gp}{... }   \PYG{c+c1}{\PYGZsh{}ax[0].legend()}
\PYG{g+gp}{... }   \PYG{n}{ax}\PYG{p}{[}\PYG{l+m+mi}{1}\PYG{p}{]}\PYG{o}{.}\PYG{n}{set\PYGZus{}xscale}\PYG{p}{(}\PYG{l+s+s1}{\PYGZsq{}}\PYG{l+s+s1}{log}\PYG{l+s+s1}{\PYGZsq{}}\PYG{p}{)}
\PYG{g+gp}{... }   \PYG{n}{ax}\PYG{p}{[}\PYG{l+m+mi}{1}\PYG{p}{]}\PYG{o}{.}\PYG{n}{set\PYGZus{}yscale}\PYG{p}{(}\PYG{l+s+s1}{\PYGZsq{}}\PYG{l+s+s1}{log}\PYG{l+s+s1}{\PYGZsq{}}\PYG{p}{)}
\PYG{g+gp}{... }   \PYG{n}{ax}\PYG{p}{[}\PYG{l+m+mi}{1}\PYG{p}{]}\PYG{o}{.}\PYG{n}{set\PYGZus{}ylabel}\PYG{p}{(}\PYG{l+s+sa}{r}\PYG{l+s+s1}{\PYGZsq{}}\PYG{l+s+s1}{\PYGZdl{}}\PYG{l+s+s1}{\PYGZbs{}}\PYG{l+s+s1}{rho }\PYG{l+s+s1}{\PYGZbs{}}\PYG{l+s+s1}{, }\PYG{l+s+s1}{\PYGZbs{}}\PYG{l+s+s1}{left[ }\PYG{l+s+s1}{\PYGZbs{}}\PYG{l+s+s1}{rm}\PYG{l+s+si}{\PYGZob{}kg\PYGZcb{}}\PYG{l+s+s1}{/}\PYG{l+s+s1}{\PYGZbs{}}\PYG{l+s+s1}{rm}\PYG{l+s+si}{\PYGZob{}m\PYGZcb{}}\PYG{l+s+s1}{\PYGZca{}3 }\PYG{l+s+s1}{\PYGZbs{}}\PYG{l+s+s1}{right]\PYGZdl{}}\PYG{l+s+s1}{\PYGZsq{}}\PYG{p}{)}
\PYG{g+gp}{... }   \PYG{n}{ax}\PYG{p}{[}\PYG{l+m+mi}{1}\PYG{p}{]}\PYG{o}{.}\PYG{n}{set\PYGZus{}xlabel}\PYG{p}{(}\PYG{l+s+sa}{r}\PYG{l+s+s1}{\PYGZsq{}}\PYG{l+s+s1}{\PYGZdl{}a\PYGZdl{}}\PYG{l+s+s1}{\PYGZsq{}}\PYG{p}{)}
\PYG{g+gp}{... }   \PYG{n}{ax}\PYG{p}{[}\PYG{l+m+mi}{1}\PYG{p}{]}\PYG{o}{.}\PYG{n}{grid}\PYG{p}{(}\PYG{n}{which}\PYG{o}{=}\PYG{l+s+s1}{\PYGZsq{}}\PYG{l+s+s1}{both}\PYG{l+s+s1}{\PYGZsq{}}\PYG{p}{,} \PYG{n}{linestyle}\PYG{o}{=}\PYG{l+s+s1}{\PYGZsq{}}\PYG{l+s+s1}{:}\PYG{l+s+s1}{\PYGZsq{}}\PYG{p}{)}
\PYG{g+gp}{... }   \PYG{n}{ax}\PYG{p}{[}\PYG{l+m+mi}{1}\PYG{p}{]}\PYG{o}{.}\PYG{n}{legend}\PYG{p}{(}\PYG{n}{fontsize}\PYG{o}{=}\PYG{l+m+mi}{14}\PYG{p}{)}
\PYG{g+gp}{... }   \PYG{n}{fig}\PYG{o}{.}\PYG{n}{tight\PYGZus{}layout}\PYG{p}{(}\PYG{p}{)}

\PYG{g+gp}{\PYGZgt{}\PYGZgt{}\PYGZgt{} }\PYG{n}{show\PYGZus{}densities}\PYG{p}{(}\PYG{n}{LCDM}\PYG{p}{,} \PYG{n}{accepts\PYGZus{}default}\PYG{o}{=}\PYG{k+kc}{True}\PYG{p}{)}
\end{sphinxVerbatim}

\noindent\sphinxincludegraphics{{cosmology_module_10_0}.png}

Notice the matter-radiation equality moment at
\(a_{eq} \sim 3 \times 10^{-4}\) and the cosmological constant that
just recently came to overtake matter as the dominant component. The
\(w\text{CDM}\) (\sphinxcode{\sphinxupquote{wcdm}}) model differs from
\(\Lambda\text{CDM}\) only by the dark energy (\sphinxcode{\sphinxupquote{de}})
equation-of-state parameter (\sphinxcode{\sphinxupquote{wd}}), which although still constant can
be different from \(-1\). Note that the energy density of dark
energy is not constant now:

\fvset{hllines={, ,}}%
\begin{sphinxVerbatim}[commandchars=\\\{\}]
\PYG{g+gp}{\PYGZgt{}\PYGZgt{}\PYGZgt{} }\PYG{n}{wCDM} \PYG{o}{=} \PYG{n}{cosmo}\PYG{o}{.}\PYG{n}{CosmologicalSetup}\PYG{p}{(}
\PYG{g+gp}{... }   \PYG{l+s+s1}{\PYGZsq{}}\PYG{l+s+s1}{wcdm}\PYG{l+s+s1}{\PYGZsq{}}\PYG{p}{,}
\PYG{g+gp}{... }   \PYG{n}{optional\PYGZus{}species}\PYG{o}{=}\PYG{p}{[}\PYG{l+s+s1}{\PYGZsq{}}\PYG{l+s+s1}{baryons}\PYG{l+s+s1}{\PYGZsq{}}\PYG{p}{,} \PYG{l+s+s1}{\PYGZsq{}}\PYG{l+s+s1}{radiation}\PYG{l+s+s1}{\PYGZsq{}}\PYG{p}{]}\PYG{p}{,}
\PYG{g+gp}{... }   \PYG{n}{derived}\PYG{o}{=}\PYG{l+s+s1}{\PYGZsq{}}\PYG{l+s+s1}{de}\PYG{l+s+s1}{\PYGZsq{}}
\PYG{g+gp}{... }\PYG{p}{)}

\PYG{g+gp}{\PYGZgt{}\PYGZgt{}\PYGZgt{} }\PYG{n}{parameters} \PYG{o}{=} \PYG{p}{\PYGZob{}}
\PYG{g+gp}{... }    \PYG{l+s+s1}{\PYGZsq{}}\PYG{l+s+s1}{h}\PYG{l+s+s1}{\PYGZsq{}}\PYG{p}{:} \PYG{l+m+mf}{0.7}\PYG{p}{,}
\PYG{g+gp}{... }    \PYG{l+s+s1}{\PYGZsq{}}\PYG{l+s+s1}{Och2}\PYG{l+s+s1}{\PYGZsq{}}\PYG{p}{:} \PYG{l+m+mf}{0.12}\PYG{p}{,}
\PYG{g+gp}{... }    \PYG{l+s+s1}{\PYGZsq{}}\PYG{l+s+s1}{Obh2}\PYG{l+s+s1}{\PYGZsq{}}\PYG{p}{:} \PYG{l+m+mf}{0.022}\PYG{p}{,}
\PYG{g+gp}{... }    \PYG{l+s+s1}{\PYGZsq{}}\PYG{l+s+s1}{wd}\PYG{l+s+s1}{\PYGZsq{}}\PYG{p}{:} \PYG{o}{\PYGZhy{}}\PYG{l+m+mf}{0.9}\PYG{p}{,}
\PYG{g+gp}{... }    \PYG{l+s+s1}{\PYGZsq{}}\PYG{l+s+s1}{T\PYGZus{}CMB}\PYG{l+s+s1}{\PYGZsq{}}\PYG{p}{:} \PYG{l+m+mf}{2.725}
\PYG{g+gp}{... }\PYG{p}{\PYGZcb{}}

\PYG{g+gp}{\PYGZgt{}\PYGZgt{}\PYGZgt{} }\PYG{n}{wCDM}\PYG{o}{.}\PYG{n}{solve\PYGZus{}background}\PYG{p}{(}\PYG{n}{parameter\PYGZus{}space}\PYG{o}{=}\PYG{n}{parameters}\PYG{p}{)}
\PYG{g+gp}{\PYGZgt{}\PYGZgt{}\PYGZgt{} }\PYG{n}{show\PYGZus{}densities}\PYG{p}{(}\PYG{n}{wCDM}\PYG{p}{,} \PYG{n}{parameter\PYGZus{}space}\PYG{o}{=}\PYG{n}{parameters}\PYG{p}{)}
\end{sphinxVerbatim}

\noindent\sphinxincludegraphics{{cosmology_module_12_0}.png}

\subsection{The cosmological models}
\label{\detokenize{themodels:the-cosmological-models}}\label{\detokenize{themodels::doc}}\label{\detokenize{themodels:modelssec}}
A few models are already implemented. I give a brief description below,
with references for works that discuss some of them in detail and works that
analyzed them with this code.
The models are objects created from the \sphinxcode{\sphinxupquote{cosmic\_objects.CosmologicalSetup}} class.
This class has a generic module \sphinxcode{\sphinxupquote{solve\_background}} that calls the \sphinxcode{\sphinxupquote{Fluid}}’s
module \sphinxcode{\sphinxupquote{rho\_over\_rho0}} of each fluid to obtain the solution for their energy
densities.
When a solution cannot be obtained directly (like in some interacting models),
a fourth-order Runge-Kutta integration is done using the function
\sphinxcode{\sphinxupquote{generic\_runge\_kutta}} from \sphinxcode{\sphinxupquote{EPIC.utils}}’s \sphinxcode{\sphinxupquote{integrators}} and the fluids{}`
\sphinxcode{\sphinxupquote{drho\_da}}.
There is an intermediate function \sphinxcode{\sphinxupquote{get\_Hubble\_Friedmann}} to calculate the
Hubble rate either by just summing the energy densities, when called from the
Runge-Kutta integration, or calculating them with \sphinxcode{\sphinxupquote{rho\_over\_rho0}}.

Some new models can be introduced in the code just by editing the
\sphinxcode{\sphinxupquote{model\_recipes.ini}}, \sphinxcode{\sphinxupquote{available\_species.ini}} and (optionally)
\sphinxcode{\sphinxupquote{default\_parameter\_values.ini}} configuration files, without needing to
rebuild and install the EPIC’s package.
The format of the configuration \sphinxcode{\sphinxupquote{.ini}} files is pretty straightforward and
the containing information can serve as a guide for what needs to be defined.

\subsubsection{The \protect\(\Lambda\text{CDM}\protect\) model}
\label{\detokenize{themodels:the-model}}
When baryons and radiation are included, the solution to this cosmology will
require values for the parameters
\(\Omega_{c0}\),
\(\Omega_{b0}\),
\(T_{\text{CMB}}\),
\(H_0\),
or
\(h\),
\(\Omega_{c0} h^2\),
\(\Omega_{b0} h^2\),
\(T_{\text{CMB}}\),
and will find \(\Omega_{\Lambda} = 1 - \left( \Omega_{c0} + \Omega_{b0} + \Omega_{r0} \right)\) or
\(\Omega_{\Lambda} h^2 = h^2 - \left( \Omega_{c0} h^2 + \Omega_{b0} h^2 + \Omega_{r0} h^2 \right)\)
if \sphinxcode{\sphinxupquote{physical}}. %
\begin{footnote}[1]\sphinxAtStartFootnote
That is, assuming \sphinxcode{\sphinxupquote{derived=lambda}}, but we could also have done, for example, \sphinxcode{\sphinxupquote{physical=False, derived=matter}}, specify \(\Omega_{\Lambda}\) and the code would get \(\Omega_{m0} = 1 - \left( \Omega_{\Lambda} + \Omega_{r0} \right)\) or, still, without specifying the derived parameter and with \sphinxcode{\sphinxupquote{physical}} true, specify all the fluids’  density parameters and get \(h = \sqrt{\Omega_{c0} h^2 + \Omega_{b0} h^2 + \Omega_{r0} h^2 + \Omega_{\Lambda} h^2}\).
\end{footnote}
The radiation density parameter \(\Omega_{r0}\) is calculated according to
the CMB temperature \(T_{\text{CMB}}\), including the contribution of the
neutrinos (and antineutrinos) of the standard model.
Extending this model to allow curvature is not completely supported yet. The
Friedmann equation is
\begin{equation*}
\begin{split}\frac{H(z)}{100 \,\, \text{km s$^{-1}$ Mpc$^{-1}$}} = \sqrt{ (\Omega_{b0} h^2 + \Omega_{c0} h^2) (1+z)^3 + \Omega_{r0} h^2 (1+z)^4 + \Omega_d h^2}\end{split}
\end{equation*}
or
\begin{equation*}
\begin{split}H(z) = H_0 \sqrt{ (\Omega_{b0} + \Omega_{c0})  (1+z)^3 + \Omega_{r0} (1+z)^4 + \Omega_d},\end{split}
\end{equation*}
\(H_0\) is in units of \(\text{km s$^{-1}$ Mpc$^{-1}$}\).
This model is identified in the code by the label \sphinxcode{\sphinxupquote{lcdm}}.

\subsubsection{The \protect\(w\text{CDM}\protect\) model}
\label{\detokenize{themodels:id2}}
Identified by \sphinxcode{\sphinxupquote{wcdm}}, this is like the standard model except that the dark
energy equation of state can be any constant \(w_d\), thus having the
\(\Lambda\text{CDM}\) model as a specific case with \(w_d = -1\).
The Friedmann equation is like the above but with the dark energy contribution
multiplied by \((1+z)^{3(1+w_d)}\).

\subsubsection{The Chevallier-Polarski-Linder parametrization}
\label{\detokenize{themodels:the-chevallier-polarski-linder-parametrization}}
The CPL parametrization %
\begin{footnote}[2]\sphinxAtStartFootnote
Chevallier M. \& Polarski D., “Accelerating Universes with scaling dark matter”. International Journal of Modern Physics D 10 (2001) 213-223; Linder E. V., “Exploring the Expansion History of the Universe”. Physical Review Letters 90 (2003) 091301.
\end{footnote} of the dark energy equation of state
\begin{equation*}
\begin{split}w(a) = w_0 + w_a \left( 1-a \right)\end{split}
\end{equation*}
is also available. In this case, the dark energy contribution in the Friedmann
equation is multiplied by
\(\left(1 + z \right)^{3\left(1 + w_0 + w_a\right)} e^{-3 w_a z /\left(1 + z\right)}\)
or \(a^{-3\left(1 + w_0 + w_a\right)} e^{-3 w_a \left(1 - a\right)}\),
in terms of the scale factor.

\subsubsection{The Barboza-Alcaniz parametrization}
\label{\detokenize{themodels:the-barboza-alcaniz-parametrization}}
The Barboza-Alcaniz dark energy equation of state parametrization %
\begin{footnote}[3]\sphinxAtStartFootnote
Barboza E. M. \& Alcaniz J. S., “A parametric model for dark energy”. Physics Letters B 666 (2008) 415-419.
\end{footnote}
\begin{equation*}
\begin{split}w(z) = w_0 + w_1 \frac{z \left(1 + z\right)}{1 + z^2}\end{split}
\end{equation*}
is implemented.
This models gives a dark energy contribution in the Friedmann equation that is
multiplied by the term \(x^{3(1+w_0)} \left( x^2 - 2 x + 2 \right)^{-3 w_1/2}\), where \(x \equiv 1/a\).

\subsubsection{The Jassal-Bagla-Padmanabhan parametrization}
\label{\detokenize{themodels:the-jassal-bagla-padmanabhan-parametrization}}
Starting with version 1.4, the JBP parametrization %
\begin{footnote}[4]\sphinxAtStartFootnote
Jassal H. K., Bagla J. S., Padmanabhan T., “\sphinxstyleemphasis{WMAP} constraints on low redshift evolution of dark energy”. Monthly Notices of the Royal Astronomical Society 356 (2005) L11-L16; Jassal H. K., Bagla J. S., Padmanabhan T., “Observational constraints on low redshift evolution of dark energy: How consistent are different observations?”. Physical Review D 72 (2005) 103503.
\end{footnote} of the equation of state
\begin{equation*}
\begin{split}w(a) = w_0 + w_1 a (1-a)\end{split}
\end{equation*}
can also be used.
In this case, \(a^{-3\left(1 + w_0\right)} e^{- 3 w_1 \left[a\left(1 - a/2\right) - 1/2\right]}\)
or \((1+z)^{3\left(1+w_0\right)} e^{3 w_1 z^2/2 \left(1+z\right)^2}\)
is the
term that goes into the dark energy contribution in the Friedmann equation.

\subsubsection{Interacting Dark Energy models}
\label{\detokenize{themodels:int-models}}\label{\detokenize{themodels:interacting-dark-energy-models}}
A comprehensive review of models that consider a possible interaction between
dark energy and dark matter is given by Wang \sphinxstyleemphasis{et al.} (2016) %
\begin{footnote}[5]\sphinxAtStartFootnote
Wang B., Abdalla E., Atrio-Barandela F., Pavón D., “Dark matter and dark energy interactions: theoretial challenges, cosmological implications and observational signatures”. Reports on Progress in Physics 79 (2016) 096901.
\end{footnote}.
In interacting models, the individual conservation equations of the two dark
fluids are violated, although still preserving the total energy conservation:
\begin{equation*}
\begin{split}\dot\rho_c + 3 H \rho_c &= Q \\
\dot\rho_d + 3 H (1 + w_d) \rho_d &= -Q.\end{split}
\end{equation*}
The shape of \(Q\) is what characterizes each model. Common forms are
proportional to \(\rho_c\), to \(\rho_d\) (both supported) or to some
combination of both (not supported in this version).

To create an instance of a coupled model (\sphinxcode{\sphinxupquote{cde}}) with
\(Q \propto \rho_c\), use:

\fvset{hllines={, ,}}%
\begin{sphinxVerbatim}[commandchars=\\\{\}]
\PYG{g+gp}{\PYGZgt{}\PYGZgt{}\PYGZgt{} }\PYG{k+kn}{from} \PYG{n+nn}{EPIC}\PYG{n+nn}{.}\PYG{n+nn}{cosmology} \PYG{k}{import} \PYG{n}{cosmic\PYGZus{}objects} \PYG{k}{as} \PYG{n}{cosmo}

\PYG{g+gp}{\PYGZgt{}\PYGZgt{}\PYGZgt{} }\PYG{n}{CDE} \PYG{o}{=} \PYG{n}{cosmo}\PYG{o}{.}\PYG{n}{CosmologicalSetup}\PYG{p}{(}
\PYG{g+gp}{... }    \PYG{l+s+s1}{\PYGZsq{}}\PYG{l+s+s1}{cde}\PYG{l+s+s1}{\PYGZsq{}}\PYG{p}{,}
\PYG{g+gp}{... }    \PYG{n}{interaction\PYGZus{}setup}\PYG{o}{=}\PYG{p}{\PYGZob{}}
\PYG{g+gp}{... }        \PYG{l+s+s1}{\PYGZsq{}}\PYG{l+s+s1}{species}\PYG{l+s+s1}{\PYGZsq{}}\PYG{p}{:} \PYG{p}{[}\PYG{l+s+s1}{\PYGZsq{}}\PYG{l+s+s1}{idm}\PYG{l+s+s1}{\PYGZsq{}}\PYG{p}{,} \PYG{l+s+s1}{\PYGZsq{}}\PYG{l+s+s1}{ide}\PYG{l+s+s1}{\PYGZsq{}}\PYG{p}{]}\PYG{p}{,}
\PYG{g+gp}{... }        \PYG{l+s+s1}{\PYGZsq{}}\PYG{l+s+s1}{parameter}\PYG{l+s+s1}{\PYGZsq{}}\PYG{p}{:} \PYG{p}{\PYGZob{}}\PYG{l+s+s1}{\PYGZsq{}}\PYG{l+s+s1}{idm}\PYG{l+s+s1}{\PYGZsq{}}\PYG{p}{:} \PYG{l+s+s1}{\PYGZsq{}}\PYG{l+s+s1}{xi}\PYG{l+s+s1}{\PYGZsq{}}\PYG{p}{\PYGZcb{}}\PYG{p}{,}
\PYG{g+gp}{... }        \PYG{l+s+s1}{\PYGZsq{}}\PYG{l+s+s1}{propto\PYGZus{}other}\PYG{l+s+s1}{\PYGZsq{}}\PYG{p}{:} \PYG{p}{\PYGZob{}}\PYG{l+s+s1}{\PYGZsq{}}\PYG{l+s+s1}{ide}\PYG{l+s+s1}{\PYGZsq{}}\PYG{p}{:} \PYG{l+s+s1}{\PYGZsq{}}\PYG{l+s+s1}{idm}\PYG{l+s+s1}{\PYGZsq{}}\PYG{p}{\PYGZcb{}}\PYG{p}{,}
\PYG{g+gp}{... }        \PYG{l+s+s1}{\PYGZsq{}}\PYG{l+s+s1}{sign}\PYG{l+s+s1}{\PYGZsq{}}\PYG{p}{:} \PYG{p}{\PYGZob{}}\PYG{l+s+s1}{\PYGZsq{}}\PYG{l+s+s1}{idm}\PYG{l+s+s1}{\PYGZsq{}}\PYG{p}{:}\PYG{l+m+mi}{1}\PYG{p}{,} \PYG{l+s+s1}{\PYGZsq{}}\PYG{l+s+s1}{ide}\PYG{l+s+s1}{\PYGZsq{}}\PYG{p}{:}\PYG{o}{\PYGZhy{}}\PYG{l+m+mi}{1}\PYG{p}{\PYGZcb{}}\PYG{p}{,}
\PYG{g+gp}{... }    \PYG{p}{\PYGZcb{}}\PYG{p}{,}
\PYG{g+gp}{... }    \PYG{n}{physical}\PYG{o}{=}\PYG{k+kc}{False}\PYG{p}{,}
\PYG{g+gp}{... }    \PYG{n}{derived}\PYG{o}{=}\PYG{l+s+s1}{\PYGZsq{}}\PYG{l+s+s1}{ide}\PYG{l+s+s1}{\PYGZsq{}}
\PYG{g+gp}{... }\PYG{p}{)}
\end{sphinxVerbatim}

The mandatory species are \sphinxcode{\sphinxupquote{idm}} and \sphinxcode{\sphinxupquote{ide}}. You can add \sphinxcode{\sphinxupquote{baryons}}
in the \sphinxcode{\sphinxupquote{optional\_species}} list keyword argument, but note that
\sphinxcode{\sphinxupquote{matter}} is not available as a combined species for this model type
since dark matter is interacting with another fluid while baryons are
not. What is new here is the \sphinxcode{\sphinxupquote{interaction\_setup}} dictionary. This is
where we tell the code which \sphinxcode{\sphinxupquote{species}} are interacting (at the moment
only an energy exchange within a pair is supported), to which of them
(\sphinxcode{\sphinxupquote{idm}}) we associate the interaction \sphinxcode{\sphinxupquote{parameter}} \sphinxcode{\sphinxupquote{xi}}, indicate
the second one (\sphinxcode{\sphinxupquote{ide}}) as having an interaction term proportional to
the other (\sphinxcode{\sphinxupquote{idm}}) and specify the sign of the interaction term for
each fluid, in this case that means \(Q_c = 3 H \xi \rho_c\) and
\(Q_d = - 3 H \xi \rho_c\).

\fvset{hllines={, ,}}%
\begin{sphinxVerbatim}[commandchars=\\\{\}]
\PYG{g+gp}{\PYGZgt{}\PYGZgt{}\PYGZgt{} }\PYG{n}{parameters} \PYG{o}{=} \PYG{p}{\PYGZob{}}\PYG{l+s+s1}{\PYGZsq{}}\PYG{l+s+s1}{Oc0}\PYG{l+s+s1}{\PYGZsq{}}\PYG{p}{:} \PYG{l+m+mf}{0.3}\PYG{p}{,} \PYG{l+s+s1}{\PYGZsq{}}\PYG{l+s+s1}{H0}\PYG{l+s+s1}{\PYGZsq{}}\PYG{p}{:} \PYG{l+m+mi}{68}\PYG{p}{,} \PYG{l+s+s1}{\PYGZsq{}}\PYG{l+s+s1}{xi}\PYG{l+s+s1}{\PYGZsq{}}\PYG{p}{:} \PYG{l+m+mf}{0.4}\PYG{p}{,} \PYG{l+s+s1}{\PYGZsq{}}\PYG{l+s+s1}{wd}\PYG{l+s+s1}{\PYGZsq{}}\PYG{p}{:} \PYG{o}{\PYGZhy{}}\PYG{l+m+mi}{1}\PYG{p}{\PYGZcb{}}
\PYG{g+gp}{\PYGZgt{}\PYGZgt{}\PYGZgt{} }\PYG{n}{CDE}\PYG{o}{.}\PYG{n}{solve\PYGZus{}background}\PYG{p}{(}\PYG{n}{parameter\PYGZus{}space}\PYG{o}{=}\PYG{n}{parameters}\PYG{p}{)}
\PYG{g+gp}{\PYGZgt{}\PYGZgt{}\PYGZgt{} }\PYG{n}{show\PYGZus{}densities}\PYG{p}{(}\PYG{n}{CDE}\PYG{p}{,} \PYG{n}{parameter\PYGZus{}space}\PYG{o}{=}\PYG{n}{parameters}\PYG{p}{)}
\end{sphinxVerbatim}

\noindent\sphinxincludegraphics{{ide_3_0}.png}

Here, I am exaggerating the value of the interaction parameter so we can
see a variation on the dark energy density that is due to the
interaction, not the equation-of-state parameter, which is \(-1\).
This same cosmology can be realized with the model type \sphinxcode{\sphinxupquote{cde\_lambda}}
without specifying the parameter \sphinxcode{\sphinxupquote{wd}}, since the \sphinxcode{\sphinxupquote{ilambda}} fluid has
fixed \(w_d = -1\). The dark matter interacting term \(Q_c\) is
positive with \(\xi\) positive, thus the lowering of the dark energy
density as its energy flows towards dark matter.

\subsubsection{Fast-varying dark energy equation-of-state models}
\label{\detokenize{themodels:fast-varying-dark-energy-equation-of-state-models}}
Models of dark energy with fast-varying equation-of-state parameter have been
studied in some works %
\begin{footnote}[6]\sphinxAtStartFootnote
Corasaniti P. S. \& Copeland E. J., “Constraining the quintessence equation of state with SnIa data and CMB peaks”. Physical Review D 65 (2002) 043004; Basset B. A., Kunz M., Silk J., “A late-time transition in the cosmic dark energy?”. Monthly Notices of the Royal Astronomical Society 336 (2002) 1217-1222; De Felice A., Nesseris S., Tsujikawa S., “Observational constraints on dark energy with a fast varying equation of state”. Journal of Cosmology and Astroparticle Physics 1205, 029 (2012).
\end{footnote}. Three such models were implemented as
described in Marcondes and Pan (2017) %
\begin{footnote}[7]\sphinxAtStartFootnote
Marcondes R. J. F. \& Pan S., “Cosmic chronometer constraints on some fast-varying dark energy equations of state”. arXiv:1711.06157 {[}astro-ph.CO{]}.
\end{footnote}. We used this code
in that work.
They have all the density parameters present in the \(\Lambda\text{CDM}\)
model besides the dark energy parameters that we describe in the following.

\paragraph{The Linder-Huterer parametrization (Model 1)}
\label{\detokenize{themodels:the-linder-huterer-parametrization-model-1}}
The model \sphinxcode{\sphinxupquote{lh}} has the free parameters
\(w_p\),
\(w_f\),
\(a_t\) and
\(\tau\) characterizing the equation of state %
\begin{footnote}[8]\sphinxAtStartFootnote
Linder E. V. \& Huterer D., “How many dark energy parameters?”. Physical Review D 72 (2005) 043509.
\end{footnote}
\begin{equation*}
\begin{split}w_d(a) = w_f + \frac{w_p - w_f}{1 + (a/a_t)^{1/\tau}}.\end{split}
\end{equation*}
\(w_p\) and \(w_f\) are the asymptotic values of \(w_d\) in the
past (\(a \to 0\)) and in the future (\(a \to \infty\)), respectively;
\(a_t\) is the scale factor at the transition epoch and \(\tau\) is the
transition width.
The Friedmann equation is
\begin{equation*}
\begin{split}\frac{H(a)^2}{H_0^2} = \frac{\Omega_{r0}}{a^4} + \frac{\Omega_{m0}}{a^3} + \frac{\Omega_{d0}}{a^{3(1+w_p)}} f_1(a),\end{split}
\end{equation*}
where
\begin{equation*}
\begin{split}f_1(a) = \left( \frac{a^{1/\tau} + a_t^{1/\tau}}{1 + a_t^{1/\tau}} \right)^{3\tau(w_p - w_f)}.\end{split}
\end{equation*}

\paragraph{The Felice-Nesseris-Tsujikawa parametrization (Model 2)}
\label{\detokenize{themodels:the-felice-nesseris-tsujikawa-parametrization-model-2}}
This FNT model \sphinxcode{\sphinxupquote{fv2}} alters the previous model to allow the dark energy
to feature an extremum value of the equation of state: \sphinxfootnotemark[6]
\begin{equation*}
\begin{split}w_d(a) = w_p + (w_0 - w_p) \, a \, \frac{1 - (a/a_t)^{1/\tau}}{1 - (1/a_t)^{1/\tau}},\end{split}
\end{equation*}
where \(w_0\) is the current value of the equation of state and the other
parameters have the interpretation as in the previous model.
The Friedmann equation is
\begin{equation*}
\begin{split}\frac{H(a)^2}{H_0^2} = \frac{\Omega_{r0}}{a^4} + \frac{\Omega_{m0}}{a^3} + \frac{\Omega_{d0}}{a^{3(1+w_p)}} e^{f_2(a)},\end{split}
\end{equation*}
with
\begin{equation*}
\begin{split}f_2(a) = 3 (w_0 - w_p) \frac{1+ (1- a_t^{-1/\tau})\tau + a \bigl[\bigl\lbrace (a/a_t)^{1/\tau} - 1 \bigr\rbrace \tau - 1 \bigr]}{(1+\tau)(1 - a_t^{-1/\tau})}.\end{split}
\end{equation*}

\paragraph{The Felice-Nesseris-Tsujikawa parametrization (Model 3)}
\label{\detokenize{themodels:the-felice-nesseris-tsujikawa-parametrization-model-3}}
Finally, we have another FNT model \sphinxcode{\sphinxupquote{fv3}} with the same parameters as in Model 2 but with equation of state \sphinxfootnotemark[6]
\begin{equation*}
\begin{split}w_d(a) = w_p + (w_0 - w_p) \, a^{1/\tau} \, \frac{1 - (a/a_t)^{1/\tau}}{1 - (1/a_t)^{1/\tau}}.\end{split}
\end{equation*}
It has a Friedmann equation identical to Model 2’s except that \(f_2(a)\) is replaced by
\begin{equation*}
\begin{split}f_3(a) = 3(w_0 - w_p) \tau \frac{2 - a_t^{-1/\tau} + a_t^{1/\tau} \bigl[(a/a_t)^{1/\tau} - 2 \bigr]}{2 \bigl(1 - a_t^{-1/\tau}\bigr)}.\end{split}
\end{equation*}

\subsection{The datasets}
\label{\detokenize{thedata:the-datasets}}\label{\detokenize{thedata::doc}}\label{\detokenize{thedata:datasec}}
Observations available for comparison with model predictions are
registered in the
\sphinxcode{\sphinxupquote{EPIC/cosmology/observational\_data/available\_observables.ini}}. They
are separated in sections (\sphinxcode{\sphinxupquote{Hz}}, \sphinxcode{\sphinxupquote{H0}}, \sphinxcode{\sphinxupquote{SNeIa}}, \sphinxcode{\sphinxupquote{BAO}} and
\sphinxcode{\sphinxupquote{CMB}}), which contain the name of the \sphinxcode{\sphinxupquote{CosmologicalSetup}}’s module for the
observable theoretical calculation (\sphinxcode{\sphinxupquote{predicting\_function\_name}}).
Each dataset goes below one of the sections, with the folder or text file
containing the data indicated.
The path is relative to the \sphinxcode{\sphinxupquote{EPIC/cosmology/observational\_data/}} folder.
If the folder name is the same as the observable
label it can be ommited. Besides, the \sphinxcode{\sphinxupquote{Dataset}} subclasses are defined
at the beginning of the \sphinxcode{\sphinxupquote{.ini}} file. Each of these classes has its own
methods for initialization and likelihood evaluation.

We choose the datasets by passing a dictionary with observables as keys and
datasets as values to the function \sphinxcode{\sphinxupquote{choose\_from\_datasets}} from
\sphinxcode{\sphinxupquote{observations}} in \sphinxcode{\sphinxupquote{EPIC.cosmology}}:

\fvset{hllines={, ,}}%
\begin{sphinxVerbatim}[commandchars=\\\{\}]
\PYG{g+gp}{\PYGZgt{}\PYGZgt{}\PYGZgt{} }\PYG{k+kn}{from} \PYG{n+nn}{EPIC}\PYG{n+nn}{.}\PYG{n+nn}{cosmology} \PYG{k}{import} \PYG{n}{observations}

\PYG{g+gp}{\PYGZgt{}\PYGZgt{}\PYGZgt{} }\PYG{n}{datasets} \PYG{o}{=} \PYG{n}{observations}\PYG{o}{.}\PYG{n}{choose\PYGZus{}from\PYGZus{}datasets}\PYG{p}{(}\PYG{p}{\PYGZob{}}
\PYG{g+gp}{... }    \PYG{l+s+s1}{\PYGZsq{}}\PYG{l+s+s1}{Hz}\PYG{l+s+s1}{\PYGZsq{}}\PYG{p}{:} \PYG{l+s+s1}{\PYGZsq{}}\PYG{l+s+s1}{cosmic\PYGZus{}chronometers}\PYG{l+s+s1}{\PYGZsq{}}\PYG{p}{,}
\PYG{g+gp}{... }    \PYG{l+s+s1}{\PYGZsq{}}\PYG{l+s+s1}{H0}\PYG{l+s+s1}{\PYGZsq{}}\PYG{p}{:} \PYG{l+s+s1}{\PYGZsq{}}\PYG{l+s+s1}{HST\PYGZus{}local\PYGZus{}H0}\PYG{l+s+s1}{\PYGZsq{}}\PYG{p}{,}
\PYG{g+gp}{... }    \PYG{l+s+s1}{\PYGZsq{}}\PYG{l+s+s1}{SNeIa}\PYG{l+s+s1}{\PYGZsq{}}\PYG{p}{:} \PYG{l+s+s1}{\PYGZsq{}}\PYG{l+s+s1}{JLA\PYGZus{}simplified}\PYG{l+s+s1}{\PYGZsq{}}\PYG{p}{,}
\PYG{g+gp}{... }    \PYG{l+s+s1}{\PYGZsq{}}\PYG{l+s+s1}{BAO}\PYG{l+s+s1}{\PYGZsq{}}\PYG{p}{:} \PYG{p}{[}
\PYG{g+gp}{... }        \PYG{l+s+s1}{\PYGZsq{}}\PYG{l+s+s1}{6dF+SDSS\PYGZus{}MGS}\PYG{l+s+s1}{\PYGZsq{}}\PYG{p}{,} \PYG{c+c1}{\PYGZsh{} z = 0.122}
\PYG{g+gp}{... }        \PYG{l+s+s1}{\PYGZsq{}}\PYG{l+s+s1}{SDSS\PYGZus{}BOSS\PYGZus{}CMASS}\PYG{l+s+s1}{\PYGZsq{}}\PYG{p}{,} \PYG{c+c1}{\PYGZsh{} z = 0.57}
\PYG{g+gp}{... }        \PYG{l+s+s1}{\PYGZsq{}}\PYG{l+s+s1}{SDSS\PYGZus{}BOSS\PYGZus{}LOWZ}\PYG{l+s+s1}{\PYGZsq{}}\PYG{p}{,} \PYG{c+c1}{\PYGZsh{} z = 0.32, 0.57}
\PYG{g+gp}{... }        \PYG{l+s+s1}{\PYGZsq{}}\PYG{l+s+s1}{SDSS\PYGZus{}BOSS\PYGZus{}QuasarLyman}\PYG{l+s+s1}{\PYGZsq{}}\PYG{p}{,} \PYG{c+c1}{\PYGZsh{} z = 2.36}
\PYG{g+gp}{... }        \PYG{l+s+s1}{\PYGZsq{}}\PYG{l+s+s1}{SDSS\PYGZus{}BOSS\PYGZus{}consensus}\PYG{l+s+s1}{\PYGZsq{}}\PYG{p}{,} \PYG{c+c1}{\PYGZsh{} z = [0.38, 0.51, 0.61]}
\PYG{g+gp}{... }        \PYG{l+s+s1}{\PYGZsq{}}\PYG{l+s+s1}{SDSS\PYGZus{}BOSS\PYGZus{}Lyalpha\PYGZhy{}Forests}\PYG{l+s+s1}{\PYGZsq{}}\PYG{p}{,} \PYG{c+c1}{\PYGZsh{} z = 2.33}
\PYG{g+gp}{... }        \PYG{c+c1}{\PYGZsh{}\PYGZsq{}WiggleZ\PYGZsq{}, \PYGZsh{} z = [0.44, 0.6, 0.73]}
\PYG{g+gp}{... }    \PYG{p}{]}\PYG{p}{,}
\PYG{g+gp}{... }    \PYG{l+s+s1}{\PYGZsq{}}\PYG{l+s+s1}{CMB}\PYG{l+s+s1}{\PYGZsq{}}\PYG{p}{:} \PYG{l+s+s1}{\PYGZsq{}}\PYG{l+s+s1}{Planck2015\PYGZus{}distances\PYGZus{}LCDM}\PYG{l+s+s1}{\PYGZsq{}}\PYG{p}{,}
\PYG{g+gp}{... }\PYG{p}{\PYGZcb{}}\PYG{p}{)}
\end{sphinxVerbatim}

The \sphinxcode{\sphinxupquote{WiggleZ}} dataset is available but is commented out because it is
correlated with the SDSS datasets and thus not supposed to be used together
with them.
Refer to the papers published by the authors of the observations for details.
Other incompatible combinations are defined in the
\sphinxcode{\sphinxupquote{conflicting\_dataset\_pairs.txt}} file in
\sphinxcode{\sphinxupquote{EPIC/cosmology/observational\_data/}}.
The code will check for these conflicts prior to proceeding with an analysis.
The returned flattened dictionary of Dataset objects will later be passed to a
MCMC analysis.
Now I describe briefly the datasets made available by the community.

\subsubsection{Type Ia supernovae}
\label{\detokenize{thedata:type-ia-supernovae}}
Two types of analyses can be made with the JLA catalogue.
One can either use the full likelihood (\sphinxcode{\sphinxupquote{JLA\_full}}) or a simplified version
based on 30 redshift bins (\sphinxcode{\sphinxupquote{JLA\_simplified}}).
Here we are using the binned data consisting of distance modulus estimates at
31 points (defining 30 bins of redshift).
If you want to use the full dataset (which makes the analysis much slower since
it involves three more nuisance parameters and requires the program to invert a
740 by 740 matrix at every iteration for the calculation of the JLA
likelihood), you need to download the covariance matrix data
(\sphinxcode{\sphinxupquote{covmat\_v6.tgz}}) from
\sphinxurl{http://supernovae.in2p3.fr/sdss\_snls\_jla/ReadMe.html}. The \sphinxcode{\sphinxupquote{covmat}}
folder must be extracted to the \sphinxcode{\sphinxupquote{jla\_likelihood\_v6}} folder.
This is not included in EPIC because the data files are too big.

Either way, the data location is the same data folder \sphinxcode{\sphinxupquote{jla\_likelihood\_v6}}.
Note that the binned dataset introduces one nuisance parameter \sphinxcode{\sphinxupquote{M}},
representing an overall shift in the absolute magnitudes, and the full dataset
introduces four nuisance parameters related to the light-curve parametrization.
See Betoule et al. (2014) %
\begin{footnote}[1]\sphinxAtStartFootnote
Betoule M. et al. “Improved cosmological constraints from a joint analysis of the SDSS-II and SNLS supernova samples”. Astronomy \& Astrophysics 568, A22 (2014).
\end{footnote} for more details.

\subsubsection{CMB distance priors}
\label{\detokenize{thedata:cmb-distance-priors}}
Constraining models with temperature or polarization anisotropy amplitudes is
not currently implemented.
However, you can include the CMB distance priors from Planck2015 %
\begin{footnote}[2]\sphinxAtStartFootnote
Huang Q.-G., Wang K., Wang S. “Distance priors from Planck 2015 data”. Journal of Cosmology and Astroparticle Physics 12 (2015) 022.
\end{footnote}
or the updated priors from Planck2018. %
\begin{footnote}[3]\sphinxAtStartFootnote
Chen L., Huang Q.\textasciitilde{}G., Wang K. “Distance Priors from Planck Final Release”. arXiv:1808.05724v1 {[}astro-ph.CO{]}.
\end{footnote}
The datasets consist of an acoustic scale \(l_A\), a shift parameter
\(R\) and the physical density of baryons \(\Omega_{b0}h^2\).
You can choose between the data for \(\Lambda\text{CDM}\) and
\(w\text{CDM}\) with either \sphinxcode{\sphinxupquote{Planck2015\_distances\_LCDM}},
\sphinxcode{\sphinxupquote{Planck2015\_distances\_wCDM}}, \sphinxcode{\sphinxupquote{Planck2018\_distances\_LCDM}}, or
\sphinxcode{\sphinxupquote{Planck2018\_distances\_wCDM}};
\sphinxcode{\sphinxupquote{Planck2015\_distances\_LCDM+Omega\_k}}
\sphinxcode{\sphinxupquote{Planck2018\_distances\_LCDM+Omega\_k}} are also available for when curvature is
supported.

\subsubsection{BAO data}
\label{\detokenize{thedata:bao-data}}
Measurements of the baryon acoustic scales from the Six Degree Field Galaxy
Survey (6dF) combined with the most recent data releases of Sloan Digital Sky
Survey (SDSS-MGS), %
\begin{footnote}[4]\sphinxAtStartFootnote
Carter P. et al. “Low Redshift Baryon Acoustic Oscillation Measurement from the Reconstructed 6-degree Field Galaxy Survey”. arXiv:1803.01746v1 {[}astro-ph.CO{]}.
\end{footnote} the LOWZ and CMASS galaxy samples of the
Baryon Oscillation Spectroscopic Survey (BOSS-LOWZ and BOSS-CMASS),
\begin{footnote}[5]\sphinxAtStartFootnote
Anderson L. et al. “The clustering of galaxies in the SDSS-III Baryon Oscillation Spectroscopic Survey: measuring \(D_A\) and \(H\) at \(z = 0.57\) from the baryon acoustic peak in the Data Release 9 spectroscopic Galaxy sample”. Monthly Notices of the Royal Astronomical Society 438 (2014) 83-101.
\end{footnote} data from
the Quasar-Lyman \(\alpha\) cross-correlation, %
\begin{footnote}[6]\sphinxAtStartFootnote
Font-Ribera A. et al. “Quasar-Lyman \(\alpha\) forest cross-correlation from BOSS DR11: Baryon Acoustic Oscillations”. Journal of Cosmology and Astroparticle Physics 05 (2014) 027.
\end{footnote}
the distribution of the Lyman \(\alpha\) forest in BOSS %
\begin{footnote}[7]\sphinxAtStartFootnote
Bautista J. E. et al. “Measurement of baryon acoustic oscillation correlations at \(z = 2.3\) with SDSS DR12 Ly \(\alpha\)-Forests”. Astronomy \& Astrophysics 603 (2017) A12.
\end{footnote}
and from the
WiggleZ Dark Energy Survey %
\begin{footnote}[8]\sphinxAtStartFootnote
Kazin E. A. et al. “The WiggleZ Dark Energy Survey: improved distance measurements to \(z = 1\) with reconstruction of the baryonic acoustic feature”. Monthly Notices of the Royal Astronomical Society 441 (2014) 3524-3542.
\end{footnote} are available in the \sphinxcode{\sphinxupquote{BAO}} folder,
as well as the latest consensus from the completed SDSS-III BOSS survey.
\begin{footnote}[9]\sphinxAtStartFootnote
Alam S. et al. “The clustering of galaxies in the completed SDSS-III Baryon Oscillation Spectroscopic Survey: cosmological analysis of the DR12 galaxy sample”. Monthly Notices of the Royal Astronomical Society 470 (2017) 2617-2652.
\end{footnote}
The observable is based on the value of
the characteristic ratio \(r_s(z_d)/D_V(z)\) between
the sound horizon \(r_s\) at decoupling time (\(z_d\)) and the
effective BAO distance \(D_V\), or some variation of this.
The respective references are given in the headers of the data files.

\subsubsection{\protect\(H(z)\protect\) data}
\label{\detokenize{thedata:data}}
These are the cosmic chronometer data.
30 measurements of the Hubble expansion rate \(H(z)\) at redshifts between
0 and 2. %
\begin{footnote}[10]\sphinxAtStartFootnote
Moresco M. et al. “A 6\% measurement of the Hubble parameter at \(z \sim 0.45\): direct evidence of the epoch of cosmic re-acceleration”. Journal of Cosmology and Astroparticle Physics 05 (2016) 014.
\end{footnote}
The values of redshift, \(H\) and the uncertainties are given in the file
\sphinxcode{\sphinxupquote{Hz/Hz\_Moresco\_et\_al\_2016.txt}}.

\subsubsection{\protect\(H_0\protect\) data}
\label{\detokenize{thedata:id11}}
The \(2.4\%\) precision local measure %
\begin{footnote}[11]\sphinxAtStartFootnote
Riess A. G. et al. “A 2.4\% determination of the local value of the Hubble constant”. The Astrophysical Journal 826 (2016) 56.
\end{footnote} of \(H_0\) is present in
\sphinxcode{\sphinxupquote{H0/local\_H0\_Riess\_et\_al\_2016.txt}}.

\subsection{The calculator GUI}
\label{\detokenize{calculator_gui::doc}}\label{\detokenize{calculator_gui:the-calculator-gui}}
Calculations like the ones presented in the previous section can also be
executed from a Graphical User Interface (GUI).
This makes it easy for the unexperienced user, although learning to use EPIC
with the interactive Python interpreter is encouraged.
The command to launch the GUI is \sphinxcode{\sphinxupquote{gui}}.
However, starting with version 1.3, you can omit this command.
Opening the GUI is the default behavior of the script.
From the terminal, activate your environment and run:

\fvset{hllines={, ,}}%
\begin{sphinxVerbatim}[commandchars=\\\{\}]
\PYGZdl{} epic.py
\end{sphinxVerbatim}

This is equivalent to \sphinxcode{\sphinxupquote{epic.py gui}} (but \sphinxcode{\sphinxupquote{python epic.py}} might be needed
depending on your system).
Optionally, you can also use \sphinxcode{\sphinxupquote{\$ python -W ignore epic.py gui}} to ignore some
warnings that will be issued by \sphinxcode{\sphinxupquote{matplotlib}} (note that this will actually
omit \sphinxstyleemphasis{all} warnings).
The following screen will appear:

\noindent\sphinxincludegraphics{{home-screen}.png}

The tab in the left is where the cosmology is specified: you can choose one of the available models implemented.
The required species will be listed. Below them, the optional fluids that the
user can add just clicking them to select/deselect.
You will always be able to choose whether or not to use physical densities.
If the cold dark matter (CDM) component in the select model does not interact
with dark energy, the option to combine CDM with baryons in a single matter
fluid will be available, in which case the optional inclusion of baryons will
be ignored.
You can then choose which fluid’s density to express in terms of the others’,
as a derived parameter, and set the interaction, if that is the case.
The energy conservation equations for the dark sector in the given model
configuration will be displayed in the box at the bottom of the frame.
Click the “Build new model” button to proceed.

\noindent\sphinxincludegraphics{{model-build}.png}

In the “Specify parameters” section, the free parameters will be displayed
together with entry fields where the use can change their values.
Adjust them to your liking.
The next section of controls allows you to choose which plots to display.
EPIC will solve the cosmology and find the background energy densities and density parameters for all fluids.
The calculation of distances over a wide range of redshifts is optional.
With this option enabled, plots of the comoving distance
\(\chi(z) = c \int_0^z \left[H(\tilde z)\right]^{-1} \mathrm{d}\tilde z\),
the angular diameter distance \(d_A\), which is equal to \(a \chi\) in
the flat universe, the luminosity distance \(d_L = d_A / a^2\), the Hubble distance \(d_H \equiv c/H(z)\) and the lookback time
\(t(z) = \int_0^z \left[\left(1+\tilde z\right)H(\tilde z) \right]^{-1} \mathrm{d}\tilde z\) are generated.
You can still customize the look of the plots with the menu buttons at the button.
They let you choose between several styles from the \sphinxcode{\sphinxupquote{matplotlib}} library,
turn on the use of \(\LaTeX\) for rendering text (which is slower), with a
few typeface options and change the size of the text labels.
These options can be changed at any moment \textendash{} existing plots will be updated with
these settings.

When you choose to include the calculation of distances, the model will be
added to the list in the bottom right.
With this, if you build other models you will be able to compare their distances
between the different model configurations.

\begin{sphinxadmonition}{note}{Note:}
If you want to add another instance of the same model (with different values of parameters) to see a comparison of distances, you need to click the “Build new model” button again, otherwise the previous model will be overwritten, since no new \sphinxcode{\sphinxupquote{CosmologicalSetup}} object is created.
\end{sphinxadmonition}

After the results for each model are presented, you can select the models that
you wish to have their distances compared and click “Compare distances”.

\noindent\sphinxincludegraphics{{model-comp-distances}.png}

The comparison is shown for all the distances listed above.
Besides the navigation controls, there are buttons in the toolbar below
the plots that enable the inclusion or removal of a bottom frame showing the
residual differences using the first model listed (i.e., amongst the selected
ones) as the reference.
In other cases, it will be possible to toggle the scale of the axes between
logarithmic and linear.
In all cases, you will be able to cycle through some options of grid, line
widths, alternate between colored lines or different line styles, omit the
legend’s title or the entire legend, and finally save the figure in \sphinxcode{\sphinxupquote{pdf}} and
\sphinxcode{\sphinxupquote{png}} formats, together with the plotted data in \sphinxcode{\sphinxupquote{txt}} files.
The size of the plots can be adjusted using the handles prior to saving.
Note that they affect only the plot currently shown.
To resize all of them at the same time, try leaving them maximized with these
controls and then resizing the app window.

\subsubsection{Calculating distances at a given redshift}
\label{\detokenize{calculator_gui:calculating-distances-at-a-given-redshift}}
After having computed the background solution for a given model, you can print
the distances at any specified redshift using the button “Calculate at”.
The results are printed in the status bar.

\noindent\sphinxincludegraphics{{distances-statusbar}.png}

\section{The MCMC module}
\label{\detokenize{MCMCmodule::doc}}\label{\detokenize{MCMCmodule:the-mcmc-module}}
This module is the original part of the program, although massively rewritten
from version 1.0.4 to 1.1.
Originally, EPIC could run with standard MCMC sampler or Parallel-Tempering MCMC.
The former has been temporarily removed to give place to a new and cleaner
implementation attempting to solve some bugs.
Use version 1.0.4 in case you need it.
The MCMC sampler comes with an experimental adaptive routine that adjusts the
covariance of the proposal multivariate Gaussian probability density for
optimal efficiency, aiming at an acceptance rate around \(0.234\).
In the following sections I briefly introduce the MCMC method and show how to
use this program to perform simulations, illustrating with examples.

Starting with version 1.3, it is also possible to perform simulations from the
EPIC’s GUI.
For long-running simulations it will still be interesting to run from the
terminal, probably on a remote machine.
But the GUI may be very useful in certain situations, when the user is too
afraind of using the command-line interface and then can run the complete
simulation from the graphical interface or at least use it to prepare the
\sphinxcode{\sphinxupquote{.ini}} file to run later from the terminal following the instructions giving
by the program.
Running the complete simulation from the GUI may be impractical if the computer
is being accessed remotely (which usually is the case when your local machine’s
processor has only a few cores) or if the simulation takes too long (using data
with tipically expensive likelihood calculations).
However, the unexperienced user may find interesting to watch how the chains
evolve until convergence, since when using the GUI intermediate plots are shown
periodically at every convergence check and the acceptance rates are printed
after every loop when the chains are written to the disk.

\subsection{Introduction to MCMC and the Bayesian method}
\label{\detokenize{intro::doc}}\label{\detokenize{intro:introduction-to-mcmc-and-the-bayesian-method}}
Users familiar with the Markov Chain Monte Carlo (MCMC) method may want to skip
to the next section.
The typical problem the user will want to tackle with this program is the
problem of parameter estimation of a given theoretical model confronted with
one or more sets of observational data.
This is a very common task in Cosmology these days, specially in the light of
numerous data from several surveys, with increasing quality.
Important discoveries are expected to be made with the data from new generation
telescopes in the next decade.

In the following I give a very brief introduction to the MCMC technique and
describe how to use program.

\subsubsection{The Bayes Theorem}
\label{\detokenize{intro:the-bayes-theorem}}
Bayesian inference is based on the inversion of the data-parameters probability
relation, which is the Bayes theorem %
\begin{footnote}[1]\sphinxAtStartFootnote
Hobson M. P., Jaffe A. H., Liddle A. R., Mukherjee P. \& Parkinson D., “Bayesian methods in cosmology”. (Cambridge University Press, 2010).
\end{footnote}.
This theorem states that the posterior
probability \(P(\theta \mid D, \mathcal{M})\) of the parameter set
\(\theta\) given the data \(D\) and other information from the model
\(\mathcal{M}\) can be given by
\begin{equation*}
\begin{split}P(\theta \mid D, \mathcal{M}) = \frac{\mathcal{L}(D \mid \theta, \mathcal{M}) \, \pi(\theta \mid \mathcal{M})}{P(D, \mathcal{M})},\end{split}
\end{equation*}
where \(\mathcal{L}(D \mid \theta, \mathcal{M})\) is the likelihood of the
data given the model parameters, \(\pi(\theta \mid \mathcal{M})\) is the
prior probability, containing any information known \sphinxstyleemphasis{a priori} about the
distribution of the parameters, and \(P(D, \mathcal{M})\) is the marginal
likelihood, also popularly known as the evidence, giving the normalization of
the posterior probability. The evidence is not required for the parameter
inference but is essential in problems of selection model, when comparing two
or more different models to see which of them is favored by the data.

Direct evaluation of \(P(\theta \mid D, \mathcal{M})\) is generally a
difficult integration in a multiparameter space that we do not know how to
perform. Usually we do know how to compute the likelihood
\(\mathcal{L}(D \mid \theta, \mathcal{M})\) that is assigned to the
experiment (most commonly a distribution that is Gaussian on the data or the
parameters),
thus the use of the Bayes theorem to give the posterior probability.
Flat priors are commonly assumed, which makes the computation of the right-hand
side of the equation above trivial.
Remember that the evidence is a normalization constant not necessary for us to
learn about the most likely values of the parameters.

\subsubsection{The Metropolis-Hastings sampler}
\label{\detokenize{intro:mh-sampler}}\label{\detokenize{intro:the-metropolis-hastings-sampler}}
The MCMC method shifts the problem of calculating the unknown posterior
probability distribution in the entire space, which can be extremly expensive
for models with large number of parameters, to the problem of sampling from the
posterior distribution.
This is possible, for example, by growing a Markov chain with new states
generated by the Metropolis sampler %
\begin{footnote}[2]\sphinxAtStartFootnote
Gayer C., “Introduction to Markov Chain Monte Carlo”. in “Handbook of Markov Chain Monte Carlo” \sphinxurl{http://www.mcmchandbook.net/}
\end{footnote}.

The Markov chain has the property that every new state depends on its current
state, and only on this current state.
Dependence on more previous states or on some statistics involving
all states is not allowed.
That can be done and can even also be useful for purposes like ours, but then
the chain can not be called Markovian.

The standard MCMC consists of generating a random state \(y\) according to
a proposal probability \(Q({} \cdot \mid x_t)\) given the current state
\(x_t\) at time \(t\).
Then a random number \(u\) is drawn from a uniform distribution between 0
and 1.
The new state is accepted if \(r \ge u\), where
\begin{equation*}
\begin{split}r = \min \left[1, \frac{P(y \mid D, \mathcal{M}) Q(x_t \mid y)}{P(x_t \mid D, \mathcal{M}) Q(y \mid x_t)} \right].\end{split}
\end{equation*}
The fraction is the Metropolis-Hastings ratio.
When the proposal function is symmetrical, \(\frac{Q(x_t \mid y)}{Q(y \mid
x_t)}\) reduces to 1 and the ratio is just the original Metropolis ratio of the
posteriors.
If the new state is accepted, we set \(x_{t+1} := y\), otherwise we repeat
the state in the chain by setting \(x_{t+1} := x_t\).

The acceptance rate \(\alpha = \frac{\text{number of accepted
states}}{\text{total number of states}}\) of a chain should be around 0.234 for
optimal efficiency %
\begin{footnote}[3]\sphinxAtStartFootnote
Roberts G. O. \& Rosenthal J. S., “Optimal scaling for various Metropolis-Hastings algorithms”. Statistical Science 16 (2001) 351-367.
\end{footnote}.
This can be obtained by tuning the parameters of the function \(Q\).
In this implementation, I use a multivariate Gaussian distribution with a
diagonal covariance matrix \(S\).

\subsubsection{The Parallel Tempering algorithm (removed in this version)}
\label{\detokenize{intro:pt-algo}}\label{\detokenize{intro:the-parallel-tempering-algorithm-removed-in-this-version}}
Standard MCMC is powerful and works in most cases but there are some problems
where the user may be better off using some other method.
Due to the characteristic behavior of a Markov chain, it is possible (and even
likely) that a chain become stuck in a single mode of a multimodal
distribution.
If two or more peaks are far away from each other, the proposal function tuned
for good performance in a peak may have difficulty escaping that peak to
explore the other, because the jump may be too short.
To overcome this inefficiency, a neat variation of MCMC, called Parallel
Tempering %
\begin{footnote}[4]\sphinxAtStartFootnote
Gregory P. C., “Bayesian logical data analysis for the physical sciences: a comparative approach with Mathematica support”. (Cambridge University Press, 2005).
\end{footnote}, favors a better exploration of the entire parameter
space in such cases thanks to an arrangement of multiple chains that are run in
parallel, each one with a different ‘’temperature’’ \(T\).
The posterior is calculated as \(\mathcal{L}^{\beta} \pi\), with
\(\beta = 1/T\).
The first chain is the one that corresponds to the real life posterior we are
interested in; the other chains, at higher temperatures, will have wider
distributions, which makes it easier to jump between peaks, thus exploring more
properly the parameter space.
Periodically, a swap of states between neighboring chains is proposed and
accepted or rejected according to a Hastings-like ratio.

\subsection{Before starting}
\label{\detokenize{beforestart::doc}}\label{\detokenize{beforestart:before-starting}}
In the section {\hyperref[\detokenize{cosmology:cosmology-module}]{\sphinxcrossref{\DUrole{std,std-ref}{The Cosmology calculator}}}} we learned how to use EPIC to set up a
cosmological model and load some datasets.
The next logical step is to calculate the probability density at a given point
of the parameter space, given that model and according to the chosen data.
This can be done as follows:

\fvset{hllines={, ,}}%
\begin{sphinxVerbatim}[commandchars=\\\{\}]
\PYG{g+gp}{\PYGZgt{}\PYGZgt{}\PYGZgt{} }\PYG{k+kn}{import} \PYG{n+nn}{EPIC}\PYG{n+nn}{.}\PYG{n+nn}{cosmology}\PYG{n+nn}{.}\PYG{n+nn}{cosmic\PYGZus{}objects} \PYG{k}{as} \PYG{n+nn}{cosmo}
\PYG{g+gp}{\PYGZgt{}\PYGZgt{}\PYGZgt{} }\PYG{k+kn}{from} \PYG{n+nn}{EPIC}\PYG{n+nn}{.}\PYG{n+nn}{cosmology} \PYG{k}{import} \PYG{n}{observations}
\PYG{g+gp}{\PYGZgt{}\PYGZgt{}\PYGZgt{} }\PYG{k+kn}{from} \PYG{n+nn}{EPIC}\PYG{n+nn}{.}\PYG{n+nn}{utils}\PYG{n+nn}{.}\PYG{n+nn}{statistics} \PYG{k}{import} \PYG{n}{Analysis}

\PYG{g+gp}{\PYGZgt{}\PYGZgt{}\PYGZgt{} }\PYG{n}{lcdm} \PYG{o}{=} \PYG{n}{cosmo}\PYG{o}{.}\PYG{n}{CosmologicalSetup}\PYG{p}{(}
\PYG{g+gp}{... }    \PYG{l+s+s1}{\PYGZsq{}}\PYG{l+s+s1}{lcdm}\PYG{l+s+s1}{\PYGZsq{}}\PYG{p}{,} \PYG{n}{physical}\PYG{o}{=}\PYG{k+kc}{False}\PYG{p}{,} \PYG{n}{derived}\PYG{o}{=}\PYG{l+s+s1}{\PYGZsq{}}\PYG{l+s+s1}{lambda}\PYG{l+s+s1}{\PYGZsq{}}
\PYG{g+gp}{... }\PYG{p}{)}

\PYG{g+gp}{\PYGZgt{}\PYGZgt{}\PYGZgt{} }\PYG{n}{datasets} \PYG{o}{=} \PYG{n}{observations}\PYG{o}{.}\PYG{n}{choose\PYGZus{}from\PYGZus{}datasets}\PYG{p}{(}\PYG{p}{\PYGZob{}}
\PYG{g+gp}{... }    \PYG{l+s+s1}{\PYGZsq{}}\PYG{l+s+s1}{Hz}\PYG{l+s+s1}{\PYGZsq{}}\PYG{p}{:} \PYG{l+s+s1}{\PYGZsq{}}\PYG{l+s+s1}{cosmic\PYGZus{}chronometers}\PYG{l+s+s1}{\PYGZsq{}}\PYG{p}{,}
\PYG{g+gp}{... }    \PYG{l+s+s1}{\PYGZsq{}}\PYG{l+s+s1}{H0}\PYG{l+s+s1}{\PYGZsq{}}\PYG{p}{:} \PYG{l+s+s1}{\PYGZsq{}}\PYG{l+s+s1}{HST\PYGZus{}local\PYGZus{}H0}\PYG{l+s+s1}{\PYGZsq{}}\PYG{p}{,}
\PYG{g+gp}{... }    \PYG{l+s+s1}{\PYGZsq{}}\PYG{l+s+s1}{SNeIa}\PYG{l+s+s1}{\PYGZsq{}}\PYG{p}{:} \PYG{l+s+s1}{\PYGZsq{}}\PYG{l+s+s1}{JLA\PYGZus{}simplified}\PYG{l+s+s1}{\PYGZsq{}}\PYG{p}{,}
\PYG{g+gp}{... }\PYG{p}{\PYGZcb{}}\PYG{p}{)}

\PYG{g+gp}{\PYGZgt{}\PYGZgt{}\PYGZgt{} }\PYG{n}{priors} \PYG{o}{=} \PYG{p}{\PYGZob{}}
\PYG{g+gp}{... }    \PYG{l+s+s1}{\PYGZsq{}}\PYG{l+s+s1}{Oc0}\PYG{l+s+s1}{\PYGZsq{}} \PYG{p}{:} \PYG{p}{[}\PYG{l+m+mi}{0}\PYG{p}{,} \PYG{l+m+mf}{0.5}\PYG{p}{]}\PYG{p}{,}
\PYG{g+gp}{... }    \PYG{l+s+s1}{\PYGZsq{}}\PYG{l+s+s1}{H0}\PYG{l+s+s1}{\PYGZsq{}} \PYG{p}{:} \PYG{p}{[}\PYG{l+m+mi}{50}\PYG{p}{,} \PYG{l+m+mi}{90}\PYG{p}{]}\PYG{p}{,}
\PYG{g+gp}{... }    \PYG{l+s+s1}{\PYGZsq{}}\PYG{l+s+s1}{M}\PYG{l+s+s1}{\PYGZsq{}} \PYG{p}{:} \PYG{p}{[}\PYG{o}{\PYGZhy{}}\PYG{l+m+mf}{0.3}\PYG{p}{,} \PYG{l+m+mf}{0.3}\PYG{p}{]}\PYG{p}{,}
\PYG{g+gp}{... }\PYG{p}{\PYGZcb{}}
\PYG{g+gp}{\PYGZgt{}\PYGZgt{}\PYGZgt{} }\PYG{n}{analysis} \PYG{o}{=} \PYG{n}{Analysis}\PYG{p}{(}\PYG{n}{datasets}\PYG{p}{,} \PYG{n}{lcdm}\PYG{p}{,} \PYG{n}{priors}\PYG{p}{)}
\PYG{g+gp}{\PYGZgt{}\PYGZgt{}\PYGZgt{} }\PYG{n}{analysis}\PYG{o}{.}\PYG{n}{log\PYGZus{}posterior}\PYG{p}{(}\PYG{n}{parameter\PYGZus{}space}\PYG{o}{=}\PYG{p}{\PYGZob{}}\PYG{l+s+s1}{\PYGZsq{}}\PYG{l+s+s1}{Oc0}\PYG{l+s+s1}{\PYGZsq{}}\PYG{p}{:} \PYG{l+m+mf}{0.3}\PYG{p}{,} \PYG{l+s+s1}{\PYGZsq{}}\PYG{l+s+s1}{H0}\PYG{l+s+s1}{\PYGZsq{}}\PYG{p}{:} \PYG{l+m+mi}{68}\PYG{p}{,} \PYG{l+s+s1}{\PYGZsq{}}\PYG{l+s+s1}{M}\PYG{l+s+s1}{\PYGZsq{}}\PYG{p}{:} \PYG{l+m+mi}{0}\PYG{p}{\PYGZcb{}}\PYG{p}{,} \PYG{n}{chi2}\PYG{o}{=}\PYG{k+kc}{True}\PYG{p}{)}
\PYG{g+go}{(\PYGZhy{}38.37946644320705, \PYGZhy{}35.38373416965306)}
\end{sphinxVerbatim}

In this example I am choosing the cosmic chronometers dataset, the Hubble constant local measurement and the simplified version of the JLA dataset.
The \sphinxcode{\sphinxupquote{Analysis}} object is created from the dictionary of datasets, the model
and a dictionary of priors in the model parameters (including nuisance
parameters related to the data).
The probability density at any point can then be calculated with the module
\sphinxcode{\sphinxupquote{log\_posterior}}, which returns the logarithm of the posterior probability
density and the logarithm of the likelihood.
Setting the option \sphinxcode{\sphinxupquote{chi2}} to \sphinxcode{\sphinxupquote{True}} (It is \sphinxcode{\sphinxupquote{False}} by default) makes the
calculation of the likelihood as \(\log \mathcal{L} = - \chi^2/2\),
dropping the usual multiplicative terms from the normalized Gaussian likelihood.
When false, the results include the contribution of the factors
\(1/\sqrt{2\pi} \sigma_i\) or the factor \(1/\sqrt{2 \pi |\textbf{C}|}\).
These are constant in most cases, making no difference to the analysis,
but in other cases, depending on the data set, the covariance matrix
\(\textbf{C}\) can depend on nuisance parameters and thus vary at each
point.

Now that we know how to calculate the posterior probability at a given point,
we can perform a Monte Carlo Markov Chain simulation to assess the confidence
regions of the model parameters.
The main script \sphinxcode{\sphinxupquote{epic.py}} accomplishes this making use of the objects and
modules here presented.

The configuration of the analysis (choice of model, datasets, priors, etc) is
defined in a \sphinxcode{\sphinxupquote{.ini}} configuration file that the program reads.
The program creates a folder in the working directory with the same name of
this \sphinxcode{\sphinxupquote{.ini}} file, if it does not already exist.
Another folder is created with the date and time for the output of each run of
the code, but you can always continue a previous run from where it stopped,
just giving the folder name instead of the \sphinxcode{\sphinxupquote{.ini}} file.
The script is stored in the \sphinxcode{\sphinxupquote{EPIC}} source folder, where the \sphinxcode{\sphinxupquote{.ini}} files
should also be placed.
The default working directory is the \sphinxcode{\sphinxupquote{EPIC}}’s parent directory, i.e., the
\sphinxcode{\sphinxupquote{epic}} repository folder.

\subsubsection{Changing the default working directory}
\label{\detokenize{beforestart:changing-the-default-working-directory}}
By default, the folders with the name of the \sphinxcode{\sphinxupquote{.ini}} files are created at the
repository root level.
But the chains can get very long and you might want to have them stored in a
different drive.
In order to set a new default location for all the new files, run:

\fvset{hllines={, ,}}%
\begin{sphinxVerbatim}[commandchars=\\\{\}]
\PYGZdl{} python define\PYGZus{}altdir.py
\end{sphinxVerbatim}

This will ask for the path of the folder where you want to save all the output
of the program and keep this information in a file \sphinxcode{\sphinxupquote{altdir.txt}}.
If you want to revert this change you can delete the \sphinxcode{\sphinxupquote{altdir.txt}} file or run
again the command above and leave the answer empty when prompted.
To change this directory temporarily you can use the argument \sphinxcode{\sphinxupquote{-{-}alt-dir}}
when running the main script.

\subsubsection{The structure of the \sphinxstyleliteralintitle{\sphinxupquote{.ini}} file}
\label{\detokenize{beforestart:the-structure-of-the-ini-file}}
Let us work with an example, with a simple flat \(\Lambda\text{CDM}\) model.
Suppose we want to constrain its parameters with \(H(z)\), supernovae data,
CMB shift parameters and BAO data.
The model parameters are the reduced Hubble constant \(h\), the present-day
values of the physical density parameters of dark matter \(\Omega_{c0} h^2\),
baryons \(\Omega_{b0} h^2\) and radiation \(\Omega_{r0} h^2\).
We will not consider perturbations, we are only constraining the parameters at
the background level.
Since we are using supernovae data we must include a nuisance parameter
\(M\), which represents a shift in the absolute magnitudes of the
supernovae.
Use of the full JLA catalogue requires the inclusion of the nuisance parameters
\(\alpha\), \(\beta\) and \(\Delta M\) from the light-curve fit.
The first section of \sphinxcode{\sphinxupquote{.ini}} is required to specify the \sphinxcode{\sphinxupquote{type}} of the model,
whether to use physical density parameters or not, and which species has the
density parameter derived from the others (e. g. from the flatness condition):

\fvset{hllines={, ,}}%
\begin{sphinxVerbatim}[commandchars=\\\{\}]
[model]
type = lcdm
physical = yes
optional species = [\PYGZsq{}baryons\PYGZsq{}, \PYGZsq{}radiation\PYGZsq{}]
derived = lambda
\end{sphinxVerbatim}

The \sphinxcode{\sphinxupquote{lcdm}} model will always have the two species \sphinxcode{\sphinxupquote{cdm}} and \sphinxcode{\sphinxupquote{lambda}}.
We are including the optional baryonic fluid and radiation, which being a
\sphinxcode{\sphinxupquote{combined species}} replaces \sphinxcode{\sphinxupquote{photons}} and \sphinxcode{\sphinxupquote{neutrinos}}.
The configurations and options available for each model are registered in the
\sphinxcode{\sphinxupquote{EPIC/cosmology/model\_recipes.ini}} file.
This section can still received the \sphinxcode{\sphinxupquote{interaction setup}} dictionary to set the
configuration of an interacting dark sector model.
Details on this are given in the previous section {\hyperref[\detokenize{themodels:int-models}]{\sphinxcrossref{\DUrole{std,std-ref}{Interacting Dark Energy models}}}}.

The second section defines the analysis: a label, datasets and specifications
about the priors ranges and distributions.
The optional property \sphinxcode{\sphinxupquote{prior distributions}} can
receive a dictionary with either \sphinxcode{\sphinxupquote{Flat}} or \sphinxcode{\sphinxupquote{Gaussian}} for each parameter.
When not specified, the code will assume flat priors by default and interpret
the list of two numbers as an interval prior range.
When \sphinxcode{\sphinxupquote{Gaussian}}, these numbers are interpreted as the parameters \(\mu\)
and \(\sigma\) of the Gaussian distribution.
In the \sphinxcode{\sphinxupquote{simulation}} section, we specify the
parameters of the diagonal covariance matrix to be used with the proposal
probability distribution in the sampler.
Values comparable to the expected standard deviation of the parameter
distributions are recommended.

\fvset{hllines={, ,}}%
\begin{sphinxVerbatim}[commandchars=\\\{\}]
[analysis]
label = \PYGZdl{}H(z)\PYGZdl{} + \PYGZdl{}H\PYGZus{}0\PYGZdl{} + SNeIa + BAO + CMB 
datasets = \PYGZob{}
    \PYGZsq{}Hz\PYGZsq{}:   \PYGZsq{}cosmic\PYGZus{}chronometers\PYGZsq{},
    \PYGZsq{}H0\PYGZsq{}:   \PYGZsq{}HST\PYGZus{}local\PYGZus{}H0\PYGZsq{},
    \PYGZsq{}SNeIa\PYGZsq{}: \PYGZsq{}JLA\PYGZus{}simplified\PYGZsq{},
    \PYGZsq{}BAO\PYGZsq{}:   [
        \PYGZsq{}6dF+SDSS\PYGZus{}MGS\PYGZsq{},
        \PYGZsq{}SDSS\PYGZus{}BOSS\PYGZus{}CMASS\PYGZsq{},
        \PYGZsq{}SDSS\PYGZus{}BOSS\PYGZus{}LOWZ\PYGZsq{},
        \PYGZsq{}SDSS\PYGZus{}BOSS\PYGZus{}QuasarLyman\PYGZsq{},
        \PYGZsq{}SDSS\PYGZus{}BOSS\PYGZus{}consensus\PYGZsq{}, 
        \PYGZsq{}SDSS\PYGZus{}BOSS\PYGZus{}Lyalpha\PYGZhy{}Forests\PYGZsq{},
        ],
    \PYGZsq{}CMB\PYGZsq{}:   \PYGZsq{}Planck2015\PYGZus{}distances\PYGZus{}LCDM\PYGZsq{},
    \PYGZcb{}
priors = \PYGZob{}
    \PYGZsq{}Och2\PYGZsq{} :  [0.08, 0.20],
    \PYGZsq{}Obh2\PYGZsq{} :  [0.02, 0.03],
    \PYGZsq{}h\PYGZsq{} : [0.5, 0.9],
    \PYGZsq{}M\PYGZsq{} : [\PYGZhy{}0.3, 0.3],
    \PYGZcb{}
prior distributions = 
fixed = \PYGZob{}
    \PYGZsq{}T\PYGZus{}CMB\PYGZsq{} : 2.7255
    \PYGZcb{}

[simulation]
proposal covariance = \PYGZob{}
    \PYGZsq{}Och2\PYGZsq{} : 1e\PYGZhy{}3,
    \PYGZsq{}Obh2\PYGZsq{} : 1e\PYGZhy{}5,
    \PYGZsq{}h\PYGZsq{} : 1e\PYGZhy{}3,
    \PYGZsq{}M\PYGZsq{}: 1e\PYGZhy{}3,
    \PYGZcb{}
\end{sphinxVerbatim}

\subsection{Running MCMC}
\label{\detokenize{MCMC:running-mcmc}}\label{\detokenize{MCMC::doc}}
This is the vanilla Monte Carlo Markov Chain with the Metropolis algorithm, as
introduced in the previous section {\hyperref[\detokenize{intro:mh-sampler}]{\sphinxcrossref{\DUrole{std,std-ref}{The Metropolis-Hastings sampler}}}}.

We will now proceed to run MCMC for the \(\Lambda\text{CDM}\) model.
The \sphinxcode{\sphinxupquote{epic.py}} script accepts five commands (besides \sphinxcode{\sphinxupquote{gui}}): \sphinxcode{\sphinxupquote{run}},
\sphinxcode{\sphinxupquote{analyze}}, \sphinxcode{\sphinxupquote{plot}}, \sphinxcode{\sphinxupquote{monitor}} and \sphinxcode{\sphinxupquote{burst}}.
The commands may be issued from any location if you are on Unix in an
environment created with \sphinxcode{\sphinxupquote{venv}}.
In other configurations you may need to run the commands from the EPIC’s home
directory and/or add \sphinxcode{\sphinxupquote{python}} at the beginning of the command.

\subsubsection{Sampling the posterior distribution}
\label{\detokenize{MCMC:sampling-the-posterior-distribution}}
Standard MCMC is the default option for sampling.
We start a MCMC simulation with the \sphinxcode{\sphinxupquote{run}} command:

\fvset{hllines={, ,}}%
\begin{sphinxVerbatim}[commandchars=\\\{\}]
\PYGZdl{} epic.py run /path/to/LCDM.ini 12 10000 \PYGZhy{}\PYGZhy{}sim\PYGZhy{}full\PYGZhy{}name MyFirstRun \PYGZhy{}\PYGZhy{}check\PYGZhy{}interval 6h
\end{sphinxVerbatim}

where the first argument is the \sphinxcode{\sphinxupquote{.ini}} file, the second is the number of
chains to be run in parallel and the third is the number of steps to be run in
each MCMC iteration.
This number of steps means that the chains will be written to the disk (the new
states are appended to the chains files) after each \sphinxcode{\sphinxupquote{steps}} states in all
chains.
A large number prevents frequent writing operations, which could impact overall
performance unnecessarily.
Each chain will create an independent process with Python’s
\sphinxcode{\sphinxupquote{multiprocessing}}, so you should not choose a
number higher than the number of CPUs \sphinxcode{\sphinxupquote{multiprocessing.cpu\_count()}} of your
machine.
The number of chains should not be less than 2.

If correlations in the initial proposal covariance matrix are needed or
desired, the user can overwrite the diagonal proposal covariance matrix with
the option \sphinxcode{\sphinxupquote{-{-}proposal-covariance}} followed by the name of the file
containing the desired covariance matrix in a table format.

The program creates a directory for this new simulation. Inside this directory,
another one named \sphinxcode{\sphinxupquote{chains}} receives the files that will store the states of
the chains, named \sphinxcode{\sphinxupquote{chain\_0.txt}}, \sphinxcode{\sphinxupquote{chain\_1.txt}}, etc.
Other relevant files are equally named but stored in the folder
\sphinxcode{\sphinxupquote{current\_states}}. These store only the last state of each chain, to allow
fast resuming from where the chain stopped, without need to loading the entire
chains.
All chains start at the same point, with coordinates given by the default
values of each parameter, unless the option \sphinxcode{\sphinxupquote{-{-}multi-start}} is given, in
which case they start at points randomly sampled from the priors.
A good starting point may help to obtain convergence faster, besides
eliminating the need for burn-in %
\begin{footnote}[1]\sphinxAtStartFootnote
When checking convergence with Gelman-Rubin method, however, burn-in is still applied.
\end{footnote}.
The name of the folder is the date-time of creation (in UTC), unless the option
\sphinxcode{\sphinxupquote{-{-}sim-full-name MyFirstRun}} is used, then \sphinxcode{\sphinxupquote{MyFirstRun}} will be the
folder name.
A custom label or tag can also be prepended with the \sphinxcode{\sphinxupquote{-{-}sim-tag my\_label}}
option, for example, the folder \sphinxcode{\sphinxupquote{my\_label-171110-173000}}.
It will be stored within another folder with the same name of the \sphinxcode{\sphinxupquote{.ini}}
file, thus in this case \sphinxcode{\sphinxupquote{LCDM/MyFirstRun}}.
The full path of the simulation will be displayed in the first line of the output.

An existing simulation can be resumed from where it last saved information to
disk with the same command, just giving the path of the simulation instead of
the \sphinxcode{\sphinxupquote{.ini}} file name, and omitting the number of chains, which has already
been defined in the first run, for example:

\fvset{hllines={, ,}}%
\begin{sphinxVerbatim}[commandchars=\\\{\}]
\PYGZdl{} epic.py run \PYGZlt{}FULL\PYGZhy{}OR\PYGZhy{}RELATIVE\PYGZhy{}PATH\PYGZhy{}TO\PYGZgt{}/LCDM/MyFirstRun/ 5000
\end{sphinxVerbatim}

Another important parameter is the tolerance accepted for convergence assessment.
By default, the program will stop when the convergence check finds, according
to the Gelman-Rubin method, convergence with \(\epsilon < 0.01\).
To change this value, you can use \sphinxcode{\sphinxupquote{-{-}tolerance 0.002}} or just \sphinxcode{\sphinxupquote{-t 2e-3}},
for example.
In the MCMC mode, the code will periodically check for convergence according
to the Gelman-Rubin method (by default it is done every two hours but can be
specified differently as \sphinxcode{\sphinxupquote{-{-}check-interval 12h}} or \sphinxcode{\sphinxupquote{-c 45min}} in the
arguments, for example.
This does not need to (and should not) be a small time interval, but the option
to specify this time in minutes or even in seconds (\sphinxcode{\sphinxupquote{30sec}}) is implemented
and available for testing purposes.

The relevant information will be displayed, in our example case looking
similar to the following:

\fvset{hllines={, ,}}%
\begin{sphinxVerbatim}[commandchars=\\\{\}]
Simulation at \PYGZlt{}FULL\PYGZhy{}PATH\PYGZhy{}TO\PYGZgt{}/LCDM/MyFirstRun
Mode MCMC.
The following datasets will be used:
    6dF+SDSS\PYGZus{}MGS
    HST\PYGZus{}local\PYGZus{}H0
    JLA\PYGZus{}simplified
    Planck2015\PYGZus{}distances\PYGZus{}LCDM
    SDSS\PYGZus{}BOSS\PYGZus{}CMASS
    SDSS\PYGZus{}BOSS\PYGZus{}LOWZ
    SDSS\PYGZus{}BOSS\PYGZus{}Lyalpha\PYGZhy{}Forests
    SDSS\PYGZus{}BOSS\PYGZus{}QuasarLyman
    SDSS\PYGZus{}BOSS\PYGZus{}consensus
    cosmic\PYGZus{}chronometers
Initiating MCMC...
\end{sphinxVerbatim}

and the MCMC will start.

The MCMC run stops if convergence is achieved with a tolerance smaller than
the default or given \sphinxcode{\sphinxupquote{tolerance}}.

The following is the output of our example after the MCMC has started.
The 10000 steps take a bit more than thirty minutes in my workstation running 12
chains in parallel.
The number of chains will not make much impact on this unless we use too many
steps by iteration and work close to the machine’s memory limit.
After approximately six hours, convergence is checked. Since it is larger than
our required \sphinxcode{\sphinxupquote{tolerance}}, the code continues with new iterations for another six
hours before checking convergence again and so on. When convergence smaller than
\sphinxcode{\sphinxupquote{tolerance}} is achieved the code makes the relevant plots and quits.

\fvset{hllines={, ,}}%
\begin{sphinxVerbatim}[commandchars=\\\{\}]
Initiating MCMC...
i 1, 10000 steps, 12 ch; 32m45s, Sat Mar 24 00:29:13 2018. Next: \PYGZti{}5h27m.
i 2, 20000 steps, 12 ch; 32m53s, Sat Mar 24 01:02:07 2018. Next: \PYGZti{}4h54m.
i 3, 30000 steps, 12 ch; 32m52s, Sat Mar 24 01:34:59 2018. Next: \PYGZti{}4h21m.
i 4, 40000 steps, 12 ch; 32m54s, Sat Mar 24 02:07:54 2018. Next: \PYGZti{}3h48m.
i 5, 50000 steps, 12 ch; 32m51s, Sat Mar 24 02:40:46 2018. Next: \PYGZti{}3h15m.
i 6, 60000 steps, 12 ch; 32m53s, Sat Mar 24 03:13:39 2018. Next: \PYGZti{}2h42m.
i 7, 70000 steps, 12 ch; 33m3s, Sat Mar 24 03:46:43 2018. Next: \PYGZti{}2h9m.
i 8, 80000 steps, 12 ch; 32m52s, Sat Mar 24 04:19:35 2018. Next: \PYGZti{}1h36m.
i 9, 90000 steps, 12 ch; 32m55s, Sat Mar 24 04:52:30 2018. Next: \PYGZti{}1h3m.
i 10, 100000 steps, 12 ch; 32m52s, Sat Mar 24 05:25:23 2018. Next: \PYGZti{}31m3s.
i 11, 110000 steps, 12 ch; 32m46s, Sat Mar 24 05:58:10 2018. Checking now...
Loading chains...                                            [\PYGZsh{}\PYGZsh{}\PYGZsh{}\PYGZsh{}\PYGZsh{}\PYGZsh{}\PYGZsh{}\PYGZsh{}\PYGZsh{}\PYGZsh{}]  12/12
Monitoring convergence...                                    [\PYGZsh{}\PYGZsh{}\PYGZsh{}\PYGZsh{}\PYGZsh{}\PYGZsh{}\PYGZsh{}\PYGZsh{}\PYGZsh{}\PYGZsh{}]  100\PYGZpc{}
R\PYGZhy{}1 tendency: 7.993e\PYGZhy{}01, 8.185e\PYGZhy{}01, 7.745e\PYGZhy{}01
i 12, 120000 steps, 12 ch; 32m49s, Sat Mar 24 06:31:21 2018. Next: \PYGZti{}5h27m.
i 13, 130000 steps, 12 ch; 32m53s, Sat Mar 24 07:04:15 2018. Next: \PYGZti{}4h54m.
i 14, 140000 steps, 12 ch; 32m49s, Sat Mar 24 07:37:04 2018. Next: \PYGZti{}4h21m.
...
\end{sphinxVerbatim}

After the first loop, the user can inspect the acceptance ratio
of the chains, updated after every loop in the section \sphinxcode{\sphinxupquote{acceptance rates}} of
the \sphinxcode{\sphinxupquote{simulation\_info.ini}} file.
Chains presenting bad performance based on this acceptance rate will be discarded.
The values considered as good performance are any rate in the interval from
\(0.1\) to \(0.5\).
This can be changed with \sphinxcode{\sphinxupquote{-{-}acceptance-limits 0.2 0.5}}, for example.
If you want to completely avoid chain removal use \sphinxcode{\sphinxupquote{-{-}acceptance-limits 0 1}}.

\subsubsection{Adaptation (beta)}
\label{\detokenize{MCMC:adaptation-beta}}
When starting a new run, use the option \sphinxcode{\sphinxupquote{-{-}adapt FREE {[}ADAPT {[}STEPS{]}{]}}}, where
\sphinxcode{\sphinxupquote{FREE}} is an integer representing the number of loops of free random-walk
before adaptation, \sphinxcode{\sphinxupquote{ADAPT}} is the actual number of loops during which
adaptation will take place, and \sphinxcode{\sphinxupquote{STEPS}} is the number of steps for both,
pre-adapting and adapting phases.
Note that the last two are optional.
If \sphinxcode{\sphinxupquote{STEPS}} is omitted, the number of steps of the regular loops will be used
(from the mandatory \sphinxcode{\sphinxupquote{steps}} argument);
if only one argument is given, it is assigned to both \sphinxcode{\sphinxupquote{FREE}} and \sphinxcode{\sphinxupquote{ADAPT}}.
At least one loop of free random-walk is necessary, so the starting states can
serve as input for the adaptation phase.
The total number of states in the two phases will be registered in the
\sphinxcode{\sphinxupquote{simulation\_info.ini}} file (\sphinxcode{\sphinxupquote{adaptive burn-in}}) as the minimal necessary
burn-in size if one wants to keep the chains Markovian.
The adaptation is an iterative process that updates the covariance matrix
\(\mathbf{S}_t\) of the proposal function for the current \(t\)-th
state based on the chains history and goes as follows.
Let \(\mathbf{\Sigma}_t\) be a matrix derived from \(\mathbf{S}_t\), defined by
\(\mathbf{\Sigma}_t \equiv \mathbf{S}_t \, m/2.38^2\), where \(m\) is
the number of free parameters.
After the initial free random-walk phase, we calculate, for a given chain, the chain mean
\begin{equation*}
\begin{split}\mathbf{\bar{x}}_t = \frac{1}{t} \sum_{i=1}^{t} \mathbf{x}_i,\end{split}
\end{equation*}
where \(\mathbf{x}_i = \left(x_{1,i}, \ldots,
x_{m,i}\right)\) is the \(i\)-th state vector of the chain.
The covariance matrix for the sampling of the next state \(t+1\) will then
be given by
\begin{equation*}
\begin{split}\mathbf{S}_{t+1} = e^{2 \theta_{t+1}} \mathbf{\Sigma}_{t+1},\end{split}
\end{equation*}
where \(\theta_{t+1} = \theta_{t} + \gamma_t \left( \eta_t - 0.234 \right)\),
\(\gamma_t = t^{-\alpha}\), \(\theta_t\) is a parameter that we set to
zero initially, \(\alpha\) is a number between \(0.5\) and \(1\),
here set to \(0.6\), \(\eta_t\) is the current acceptance rate,
targeted at \(0.234\), and
\begin{equation*}
\begin{split}\mathbf{\Sigma}_{t+1} = \left(1 - \gamma_t \right) \mathbf{\Sigma}_t + \gamma_t \, \left(\mathbf{x}_t - \mathbf{\bar{x}}_t\right)^T \left(\mathbf{x}_t - \mathbf{\bar{x}}_t\right).\end{split}
\end{equation*}
In this program, the covariance matrix is updated based on the first chain and
applied to all chains.
This method is mostly based on the adaptation applied to the parallel tempering
algorithm by Łącki \& Miasojedow (2016) %
\begin{footnote}[2]\sphinxAtStartFootnote
Łącki M. K. \& Miasojedow B. “State-dependent swap strategies and automatic reduction of number of temperatures in adaptive parallel tempering algorithm”. Statistics and Computing 26, 951\textendash{}964 (2016).
\end{footnote}.
The idea is to obtain a covariance matrix such that the asymptotic acceptance
rate is \(0.234\), considered to be a good value.
Note, however, that this feature is in beta and may require some trial and
error with the choice of both free random-walk and adaptation phases lengths.
It might take too long for the proposal covariance matrix to converge and
the simulation is not guaranteed to keep good acceptance rates with
the final matrix obtained after the adaptation phase.

\subsubsection{Analyzing the chains}
\label{\detokenize{MCMC:analyzing-the-chains}}
Once convergence is achieved and MCMC is finished, the code reads the chains
and extract the information for the parameter inference.
But this can be done to assess the results while MCMC is still going on.
The distributions are compiled for a nice plot with the command \sphinxcode{\sphinxupquote{analyze}}:

\fvset{hllines={, ,}}%
\begin{sphinxVerbatim}[commandchars=\\\{\}]
\PYGZdl{} epic.py analyze \PYGZlt{}FULL\PYGZhy{}OR\PYGZhy{}RELATIVE\PYGZhy{}PATH\PYGZhy{}TO\PYGZgt{}/LCDM/MyFirstRun/
\end{sphinxVerbatim}

Histograms are generated for the marginalized distributions using 20 bins or
any other number given with \sphinxcode{\sphinxupquote{-{-}bins}} or \sphinxcode{\sphinxupquote{-b}}.
At any time one can run this command, optionally with \sphinxcode{\sphinxupquote{-{-}convergence}} or
\sphinxcode{\sphinxupquote{-c}}  to check the state of convergence.
By default, this will calculate \(\hat R^p - 1\) for twenty
different sizes (which you can change with \sphinxcode{\sphinxupquote{-{-}GR-steps}}) considering the size
of the chains to provide an idea of the evolution of the convergence.

Convergence is assessed based on the Gelman-Rubin criterium.
I will not enter into details of the method here, but I refer the reader to the
original papers %
\begin{footnote}[4]\sphinxAtStartFootnote
Gelman A \& Rubin D. B. “Inference from Iterative Simulation Using Multiple Sequences”. Statistical Science 7 (1992) 457-472.
\end{footnote} %
\begin{footnote}[5]\sphinxAtStartFootnote
Brooks S. P. \& Gelman A. “General Methods for Monitoring Convergence of Iterative Simulations”. Journal of Computational and Graphical Statistics 7 (1998) 434.
\end{footnote} for more information.
The variation of the original method for multivariate distribution is implemented.
When using \sphinxcode{\sphinxupquote{MCMC}}, all the chains are checked for convergence and the final
resulting distribution which is analyzed and plotted is the concatenation of
all the chains, since they are essentially all the same once they have
converged.

\paragraph{Additional options}
\label{\detokenize{MCMC:additional-options}}
If for some reason you want to view the results for an intermediate point of
the simulation, you can tell the script to \sphinxcode{\sphinxupquote{-{-}stop-at 18000}}, everything will
be analyzed until that point.

If you want to check the random walk of the chains you can plot the sequences
with \sphinxcode{\sphinxupquote{-{-}plot-sequences}}. This will make a grid plot containing all the chains
and all parameters. Keep in mind that this can take some time and generate a
big output file if the chains are very long. You can contour this problem by
thinning the distributions by some factor \sphinxcode{\sphinxupquote{-{-}thin 10}}, for example.
This also applies for the calculation of the correlation of the parameters in
each chain, enabled with the \sphinxcode{\sphinxupquote{-{-}correlation-function}} option.

We generally represent distributions by their histograms but sometimes we may
prefer to exhibit smooth curves. Although it is possible to choose a higher
number of histogram bins, this may not be sufficient and may required much more
data.
Much better (although possibly slow when there are too many states) are the
kernel density estimates (KDE) tuned for Gaussian-like distributions %
\begin{footnote}[6]\sphinxAtStartFootnote
Kristan M., Leonardis A., Skocaj D. “Multivariate online kernel density estimation with Gaussian kernels”. Pattern Recognit 44 (2011) 2630-2642.
\end{footnote}.
To obtain smoothed shapes use \sphinxcode{\sphinxupquote{-{-}kde}}. This will compute KDE curves
for the marginalized parameter distributions and also the two-parameter joint
posterior probability distributions.

\subsubsection{Making the triangle plots}
\label{\detokenize{MCMC:making-the-triangle-plots}}
The previous command will also produce the plots automatically (unless
supressed with \sphinxcode{\sphinxupquote{-{-}dont-plot}}), but you can always redraw everything when you
want, maybe you would like to tweak some colors? Loading the chains again is
not necessary since the analysis already saves the information for the plots
anyway.
This is done with:

\fvset{hllines={, ,}}%
\begin{sphinxVerbatim}[commandchars=\\\{\}]
\PYGZdl{} epic.py plot \PYGZlt{}FULL\PYGZhy{}OR\PYGZhy{}RELATIVE\PYGZhy{}PATH\PYGZhy{}TO\PYGZgt{}/LCDM/MyFirstRun/
\end{sphinxVerbatim}

The code will find the longest chain analysis results and plot them.
In this plot, you can, for example, choose the main color with \sphinxcode{\sphinxupquote{-{-}color m}}
for magenta, or with any other Python color name, and omit the best-fit point mark
with \sphinxcode{\sphinxupquote{-{-}no-best-fit-marks}}.
You can always use any of \sphinxcode{\sphinxupquote{C0}}, \sphinxcode{\sphinxupquote{C1}}, …, \sphinxcode{\sphinxupquote{C9}}, which refer to the
colors in the current style’s palette.
The default option is \sphinxcode{\sphinxupquote{C0}}.

Plot ranges and eventually necessary factors of power of 10 are automatically
handled by the code, unless you use \sphinxcode{\sphinxupquote{-{-}no-auto-range}} and \sphinxcode{\sphinxupquote{-{-}no-auto-factors}}
or choose your own ranges or powers of 10 by specifying a dictionary with
parameter label as keys and a pair of floats in a list or an integer power of
10 for the entries \sphinxcode{\sphinxupquote{custom range}} and \sphinxcode{\sphinxupquote{custom factors}} under the section
\sphinxcode{\sphinxupquote{analysis}} in the \sphinxcode{\sphinxupquote{.ini}} file.

If you have used the option \sphinxcode{\sphinxupquote{-{-}kde}} previously in the analysis you need to
specify it here too to make the corresponding plots, otherwise the histograms
will be drawn.

The \(1\sigma\) and \(2\sigma\) confidence levels are shown and are
written to the file \sphinxcode{\sphinxupquote{hist\_table.tex}} (and \sphinxcode{\sphinxupquote{kde\_table.tex}} if it is the
case) inside the folder with the size of the chain preceeded by the letter
\sphinxcode{\sphinxupquote{n}}.
These \(\LaTeX\)-ready files can be compiled to make a nice table in a PDF
file or included in your paper as you want.
To view and save information for more levels %
\begin{footnote}[3]\sphinxAtStartFootnote
Choice limited up to the fifth level. Notice that high sigma levels might be difficult to find due to the finite sample sizes when the sample is not sufficiently large.
\end{footnote}, use \sphinxcode{\sphinxupquote{-{-}levels 1 2 3 4 5}}.

\paragraph{Additional options}
\label{\detokenize{MCMC:id7}}
You can further tweak your plots and tables. \sphinxcode{\sphinxupquote{-{-}use-tex}} will make the program
render the plot using \(\LaTeX\), fitting nicely to the rest of your paper.
You can plot Gaussian fits together with the histograms (the Gaussian curve and
the two first sigma levels), using \sphinxcode{\sphinxupquote{-{-}show-gaussian-fits}}, but probably only for
the sake of comparison since this is not usually done in publications.
\sphinxcode{\sphinxupquote{-{-}fmt}} can be used to set the number of figures to report the results
(default is \sphinxcode{\sphinxupquote{5}}).
There is \sphinxcode{\sphinxupquote{-{-}font-size}} to choose the size of the fonts, \sphinxcode{\sphinxupquote{-{-}show-hist}} to
plot the histograms together with the smooth curves when \sphinxcode{\sphinxupquote{-{-}kde}} is used,
\sphinxcode{\sphinxupquote{-{-}no-best-fit-marks}} to omit the indication of the
coordinates of the best-fit point, \sphinxcode{\sphinxupquote{-{-}png}} to save images in png format
besides the pdf files.

If you have nuisance parameters, you can opt to not show them with
\sphinxcode{\sphinxupquote{-{-}exclude nuisance}}.
This option can actually take the label of any parameter you choose not to
include in the plot.

Going beyond the standard model and trying to detect a new parameter? After
making kernel density estimates, you can calculate how many sigmas of detection
with the option \sphinxcode{\sphinxupquote{-{-}detect xi}}, for example, where \sphinxcode{\sphinxupquote{xi}} is the label of the
parameter in the analysis. Note that this will only work if you also include
the option \sphinxcode{\sphinxupquote{-{-}kde}} at this point.

Below we see the triangle plot of the histograms, with the default settings,
in comparison with a perfected version using the smoothed distributions, the
Python color \sphinxcode{\sphinxupquote{C9}}, the \(\LaTeX\) renderer, including the \(3\sigma\)
confidence level and excluding the nuisance parameter \(M\).

\begin{figure}[!tb]
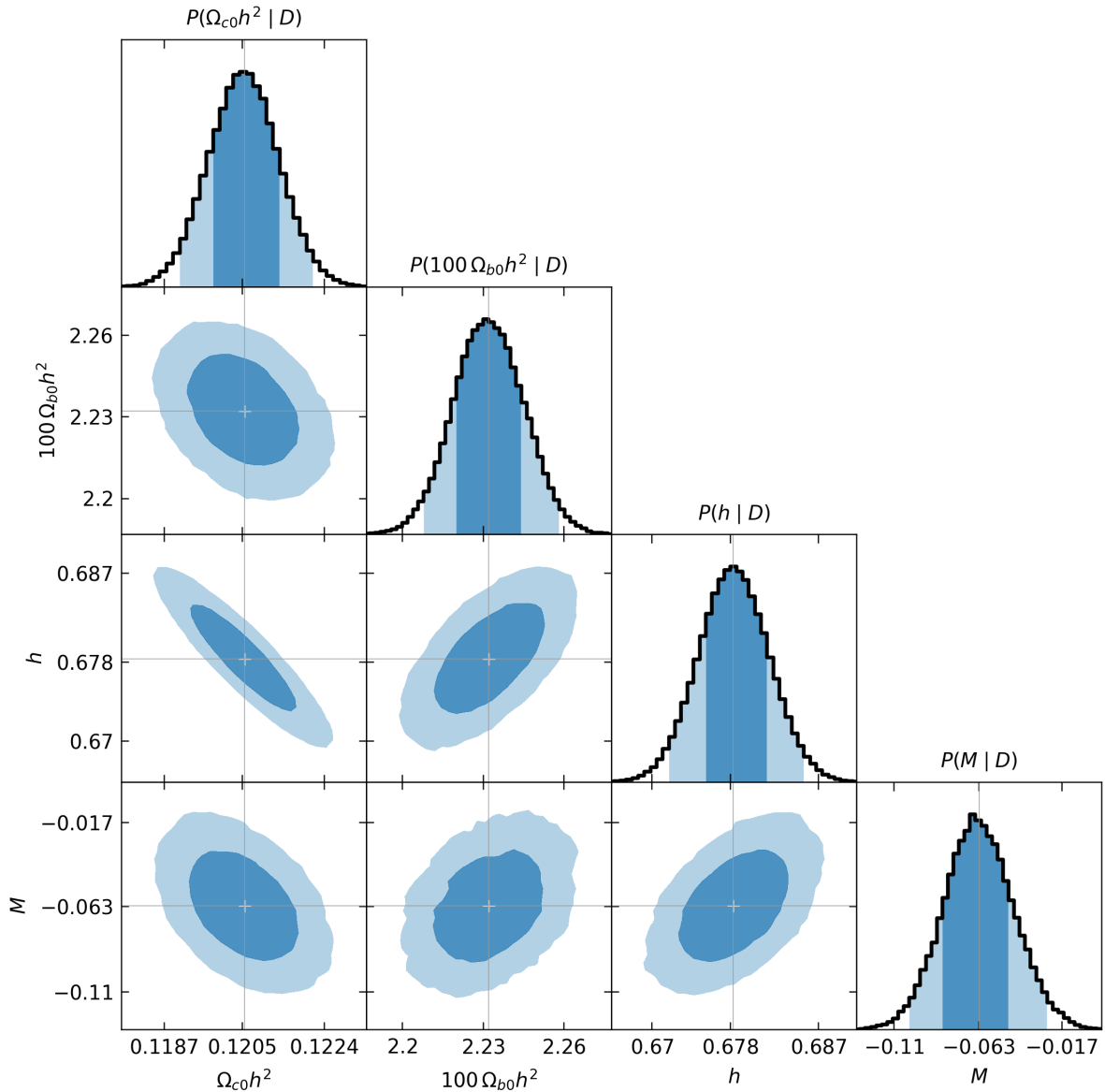

\centering
\capstart

\noindent\sphinxincludegraphics{{LCDMdefault}.png}
\caption{Triangle plot with default configurations for histograms.}\label{\detokenize{MCMC:id8}}\end{figure}

\begin{figure}[!tb]
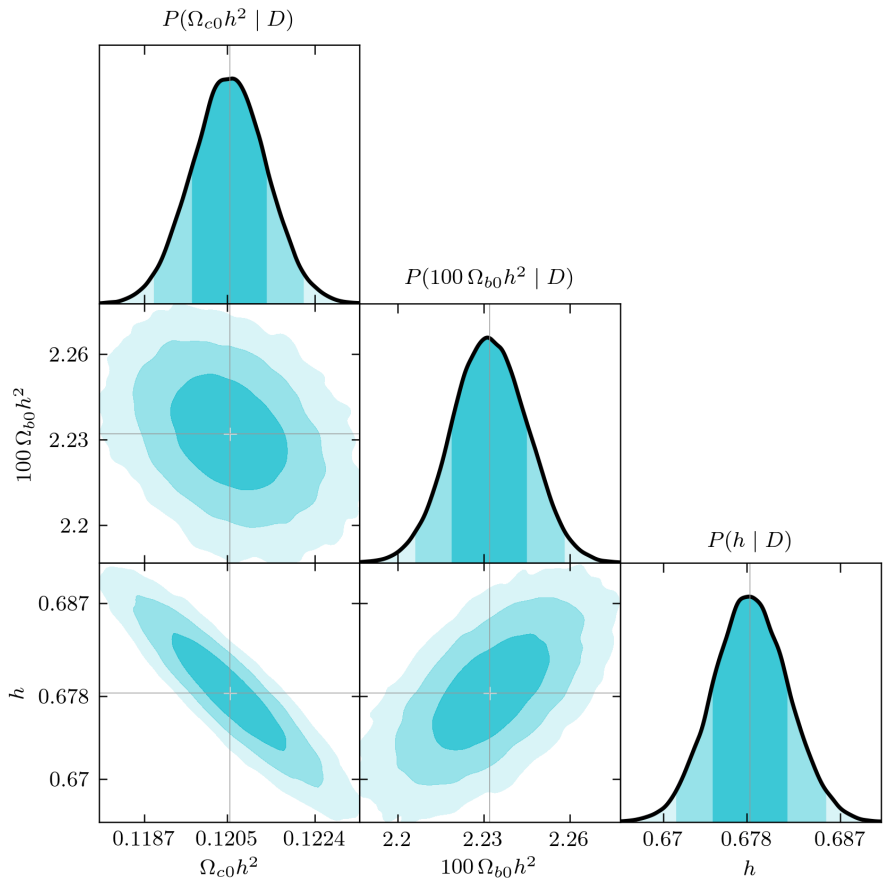

\centering
\capstart

\noindent\sphinxincludegraphics{{LCDMperfected}.png}
\caption{Customized plot with smoothed distributions.}\label{\detokenize{MCMC:id9}}\end{figure}

\paragraph{Combining two or more simulations in one plot}
\label{\detokenize{MCMC:combining-two-or-more-simulations-in-one-plot}}
You just need to run \sphinxcode{\sphinxupquote{epic.py plot}} with two or more paths in the
arguments.
I illustrate this with two simulation for the same simplified
\(\Lambda\text{CDM}\) model, with cold dark matter and \(\Lambda\)
only, one with \(H(z)\) and \(H_0\) data, the other with the same data
plus the simplified supernovae dataset from JLA.
It is then interesting to plot both realizations together so we can see the effect that including a dataset has on the results:

\fvset{hllines={, ,}}%
\begin{sphinxVerbatim}[commandchars=\\\{\}]
\PYGZdl{} epic.py plot \PYGZbs{}
\PYGZlt{}FULL\PYGZhy{}OR\PYGZhy{}RELATIVE\PYGZhy{}PATH\PYGZhy{}TO\PYGZgt{}/HLCDM+SNeIa/H\PYGZus{}and\PYGZus{}SN/ \PYGZbs{}
\PYGZlt{}FULL\PYGZhy{}OR\PYGZhy{}RELATIVE\PYGZhy{}PATH\PYGZhy{}TO\PYGZgt{}/HLCDM/H\PYGZus{}only/ \PYGZbs{}
\PYGZhy{}\PYGZhy{}kde \PYGZhy{}\PYGZhy{}use\PYGZhy{}tex \PYGZhy{}\PYGZhy{}plot\PYGZhy{}prefix comparison \PYGZhy{}\PYGZhy{}no\PYGZhy{}best\PYGZhy{}fit\PYGZhy{}marks
\end{sphinxVerbatim}

\begin{figure}[!tb]
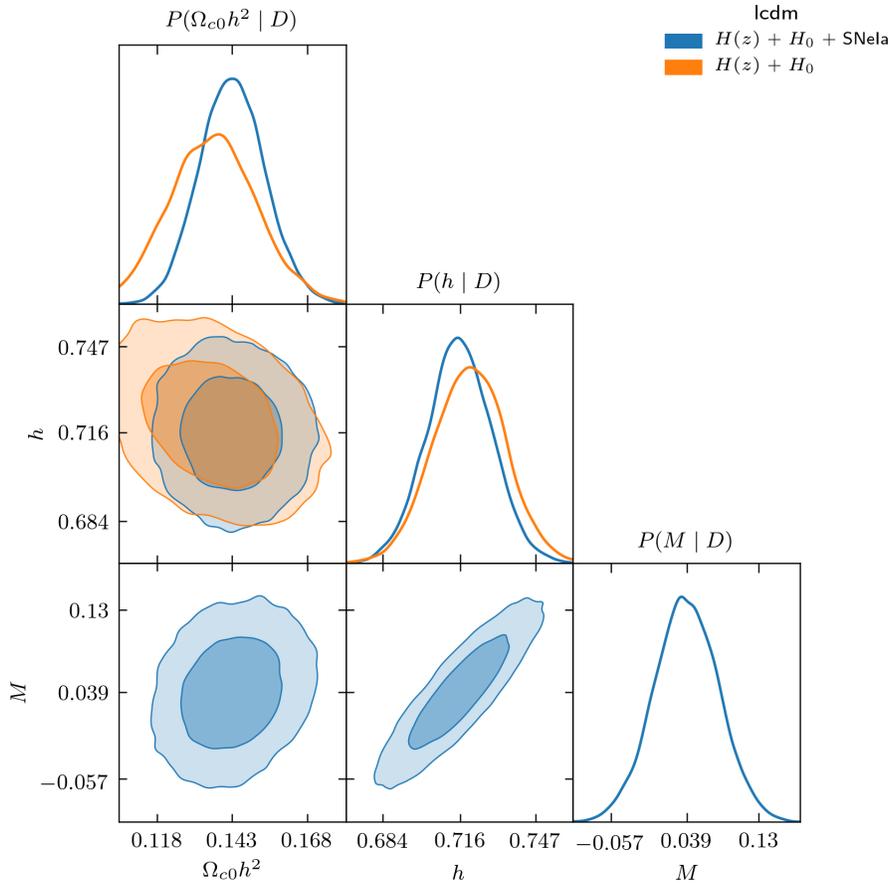

\centering
\capstart

\noindent\sphinxincludegraphics{{comparison}.png}
\caption{Results for two different analyses.}\label{\detokenize{MCMC:id10}}\end{figure}

You can combine as many results as you like.
When this is done, the list of parameters will be determined from the first
simulation given in the command by its path.
Automatic ranges for the plots are determined from the constraints of the first
simulation as well.
Since different simulations might give considerably different results
that could go outside of the ranges of one of them, consider using
\sphinxcode{\sphinxupquote{-{-}no-auto-range}} or specifying custom intervals in the \sphinxcode{\sphinxupquote{.ini}} file if
needed.
The \sphinxcode{\sphinxupquote{-{-}plot-prefix}} option specifies the prefix for the name of the pdf file
that will be generated.
If you omit this option, \sphinxcode{\sphinxupquote{grids2s}} will be used.
All other settings are optional.

A legend will be included in the top right corner using the labels defined in
the \sphinxcode{\sphinxupquote{.ini}} files, under the \sphinxcode{\sphinxupquote{analysis}} section.
The legend title uses the model name of the first simulation in the arguments.
This is intended for showing, at the same time, results from different datasets
with the same model.

When generating plots of different analyses combined, you can customize the
appearance in two ways.
The first one, is changing the \sphinxcode{\sphinxupquote{-{-}color-scheme}} option, which defaults to
\sphinxcode{\sphinxupquote{tableau}}, to either \sphinxcode{\sphinxupquote{xkcd}}, \sphinxcode{\sphinxupquote{css4}} or \sphinxcode{\sphinxupquote{base}}.
These are color palettes from \sphinxcode{\sphinxupquote{matplotlib.\_color\_data}}.
\sphinxcode{\sphinxupquote{XKCD\_COLORS}} and \sphinxcode{\sphinxupquote{CSS4\_COLORS}} are dictionaries containing 949 and 148
colors each one, respectively.
Because they are not ordered dictionaries, each time you
plot using them you will get a different color scheme with random
colors, so have fun!
The four options can be followed by options like \sphinxcode{\sphinxupquote{-light}}, \sphinxcode{\sphinxupquote{-dark}}, etc
(run \sphinxcode{\sphinxupquote{\$ epic.py plot -{-}help}} to see the full list).
These options just filter the lists returning only the colors that have that
characteristic in their names.

Another way to customize plots is to specify a style with \sphinxcode{\sphinxupquote{-{-}style}}.
The available options are \sphinxcode{\sphinxupquote{default}}, \sphinxcode{\sphinxupquote{bmh}}, \sphinxcode{\sphinxupquote{dark\_background}},
\sphinxcode{\sphinxupquote{ggplot}}, \sphinxcode{\sphinxupquote{Solarize\_Light2}}, \sphinxcode{\sphinxupquote{seaborn-bright}},  \sphinxcode{\sphinxupquote{seaborn-colorblind}},
\sphinxcode{\sphinxupquote{seaborn-dark-palette}}, \sphinxcode{\sphinxupquote{seaborn-muted}}, \sphinxcode{\sphinxupquote{seaborn-pastel}} and
\sphinxcode{\sphinxupquote{seaborn-whitegrid}}.
These styles change the line colors and some also redefine the page background,
text labels, plot background, etc.
Be aware, however, that the previous option \sphinxcode{\sphinxupquote{-{-}color-scheme}} will overwrite
the line colors of the style, unless you leave it in the default \sphinxcode{\sphinxupquote{tableau}}
option.

\paragraph{Visualizing chain sequences and convergence}
\label{\detokenize{MCMC:visualizing-chain-sequences-and-convergence}}
If you include the option \sphinxcode{\sphinxupquote{-{-}sequences}} with the \sphinxcode{\sphinxupquote{analyze}} command you will
get a plot like this

\begin{figure}[!tb]
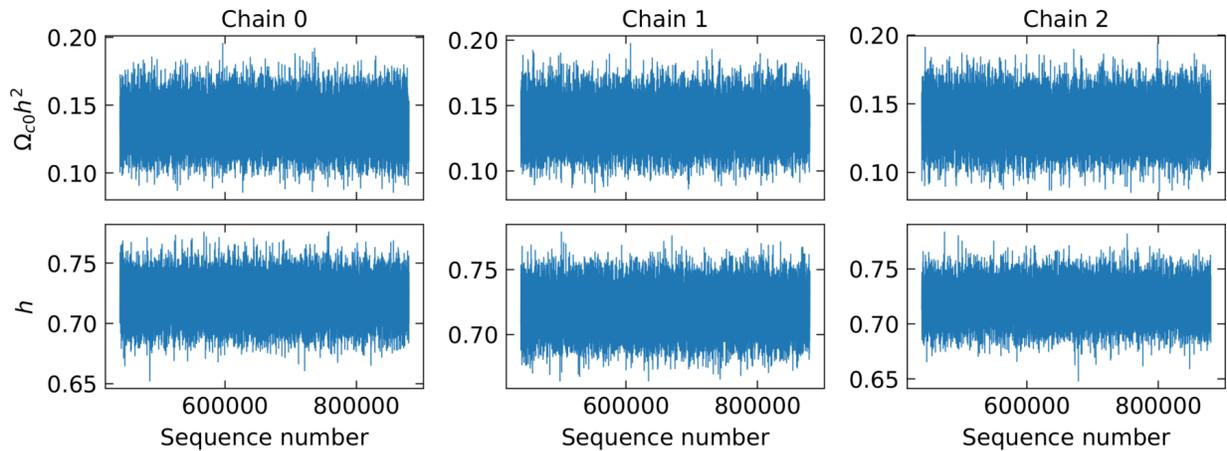

\centering
\capstart

\noindent\sphinxincludegraphics{{sequences}.png}
\caption{Chain sequences along each parameter axis (only the first few chains shown here).}\label{\detokenize{MCMC:id11}}\end{figure}

When monitoring convergence, the values of \(\hat{R}^p - 1\) at twenty (or
\sphinxcode{\sphinxupquote{-{-}GR-steps}}) different lengths for the multivariate analysis and separate
one-dimensional analyses for each parameter are plotted in the files
\sphinxcode{\sphinxupquote{monitor\_convergence\_\textless{}N\textgreater{}.pdf}} and \sphinxcode{\sphinxupquote{monitor\_each\_parameter\_\textless{}N\textgreater{}.pdf}}, where
\sphinxcode{\sphinxupquote{N}} is the total utilized length of the chains.
The absolute values of the between-chains and within-chains variances,
\(\hat{V}\) and \(W\) are also shown.
For our \(\Lambda\text{CDM}\) example, we got

\begin{figure}[!tb]
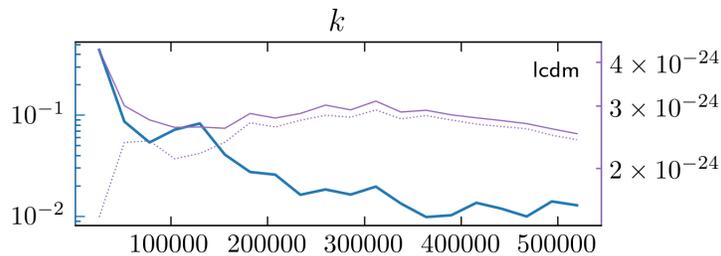

\centering
\capstart

\noindent\sphinxincludegraphics{{LCDMconvergence}.png}
\caption{Multivariate convergence analysis.}\label{\detokenize{MCMC:id12}}\end{figure}

\begin{figure}[!tb]
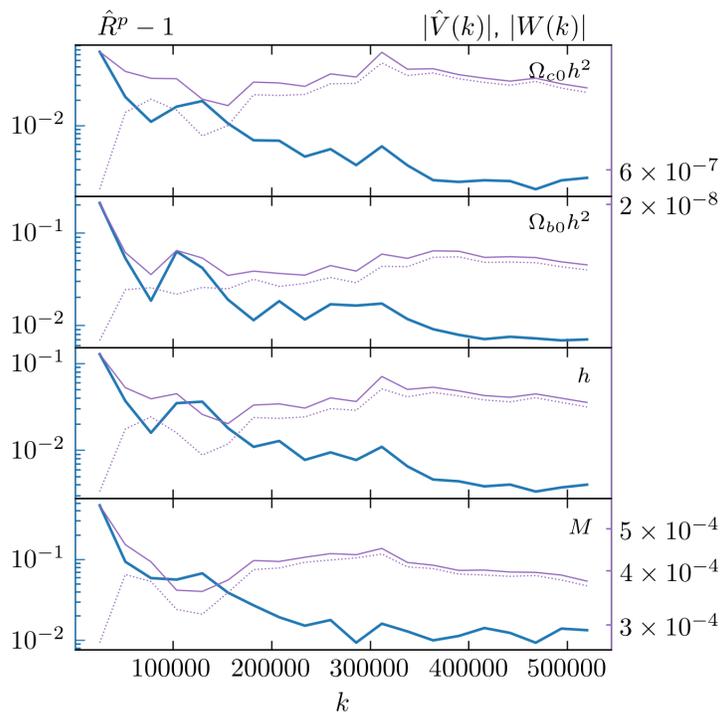

\centering
\capstart

\noindent\sphinxincludegraphics{{LCDMeachparameter}.png}
\caption{Individual parameter convergence monitoring.}\label{\detokenize{MCMC:id13}}\end{figure}

To generate these plots, one needs first run the script with the command
\sphinxcode{\sphinxupquote{analyze}} and the \sphinxcode{\sphinxupquote{-{-}convergence}} option; then, run:

\fvset{hllines={, ,}}%
\begin{sphinxVerbatim}[commandchars=\\\{\}]
\PYGZdl{} epic.py monitor \PYGZlt{}FULL\PYGZhy{}OR\PYGZhy{}RELATIVE\PYGZhy{}PATH\PYGZhy{}TO\PYGZgt{}/LCDM/MyFirstRun
\end{sphinxVerbatim}

The first argument can be multiple paths to simulations, in which case the
first plot above will have panes for each simulation, one below each other.
There are still the options \sphinxcode{\sphinxupquote{-{-}use-tex}} and \sphinxcode{\sphinxupquote{-{-}png}}, as with the \sphinxcode{\sphinxupquote{plot}}
command.
Finally, the correlation of the chains can also be inspected if we use the
option \sphinxcode{\sphinxupquote{-{-}correlation-function}}.
This includes all the cross- and auto-correlations.

\begin{figure}[!tb]
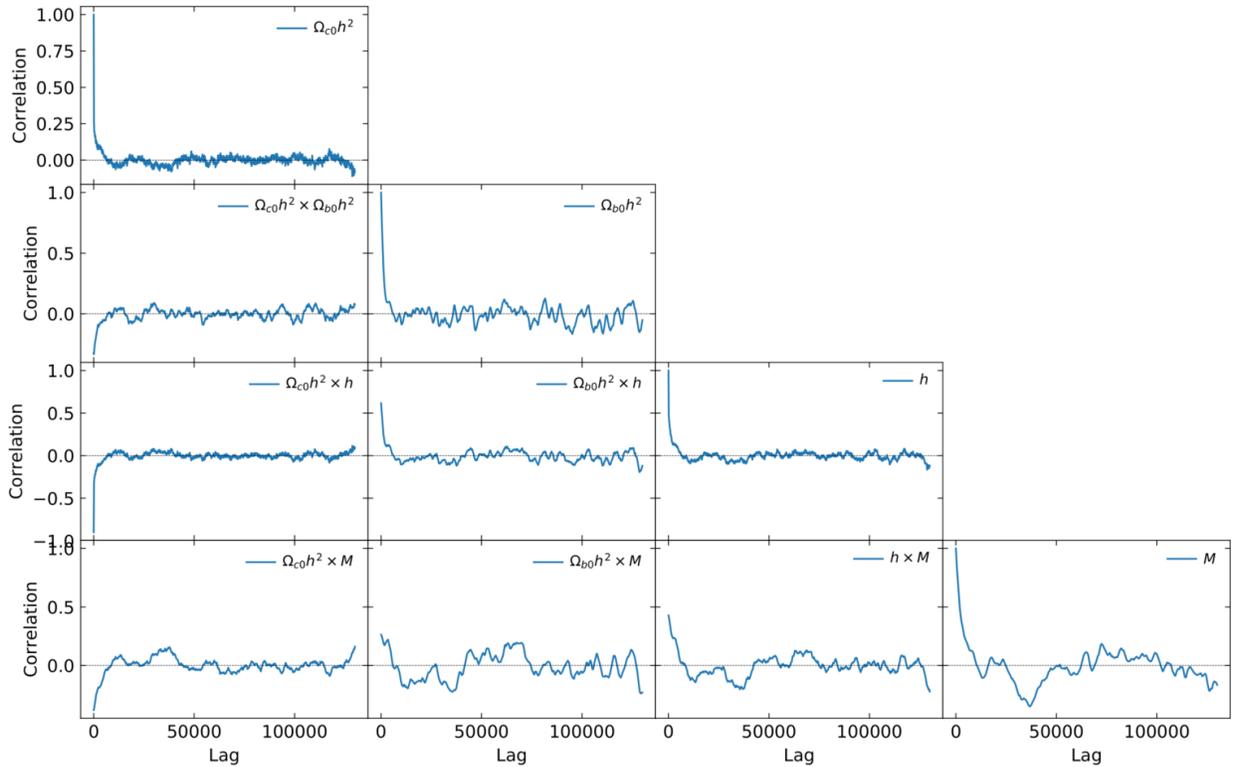

\centering
\capstart

\noindent\sphinxincludegraphics{{LCDMacf}.png}
\caption{Correlations in a chain (here using a factor \sphinxcode{\sphinxupquote{-{-}thin 15}}).}\label{\detokenize{MCMC:id14}}\end{figure}

\paragraph{Visualizing chains separately}
\label{\detokenize{MCMC:visualizing-chains-separately}}
With the command \sphinxcode{\sphinxupquote{burst}}, the chains can be viewed separately, which may be
useful to inspect unexpected behaviors of the distributions of the chains
combined.
Use:

\fvset{hllines={, ,}}%
\begin{sphinxVerbatim}[commandchars=\\\{\}]
\PYGZdl{} epic.py burst \PYGZlt{}FULL\PYGZhy{}OR\PYGZhy{}RELATIVE\PYGZhy{}PATH\PYGZhy{}TO\PYGZgt{}/\PYGZlt{}SIMULATION\PYGZgt{}
\end{sphinxVerbatim}

The script uses \sphinxcode{\sphinxupquote{analyze}} for each chain separately, via the \sphinxcode{\sphinxupquote{analyze}}’s
option \sphinxcode{\sphinxupquote{-{-}use-chain}} and then \sphinxcode{\sphinxupquote{plot}} with all the analyses performed.
The option \sphinxcode{\sphinxupquote{-{-}use-chain}} is also available for \sphinxcode{\sphinxupquote{burst}}, as well as the
options \sphinxcode{\sphinxupquote{-{-}bins}}, \sphinxcode{\sphinxupquote{-{-}kde}}, \sphinxcode{\sphinxupquote{-{-}use-tex}}, \sphinxcode{\sphinxupquote{-{-}png}} and \sphinxcode{\sphinxupquote{-{-}plot-prefix}}.
The figure bellow illustrates the result of this command.

\begin{figure}[!tb]
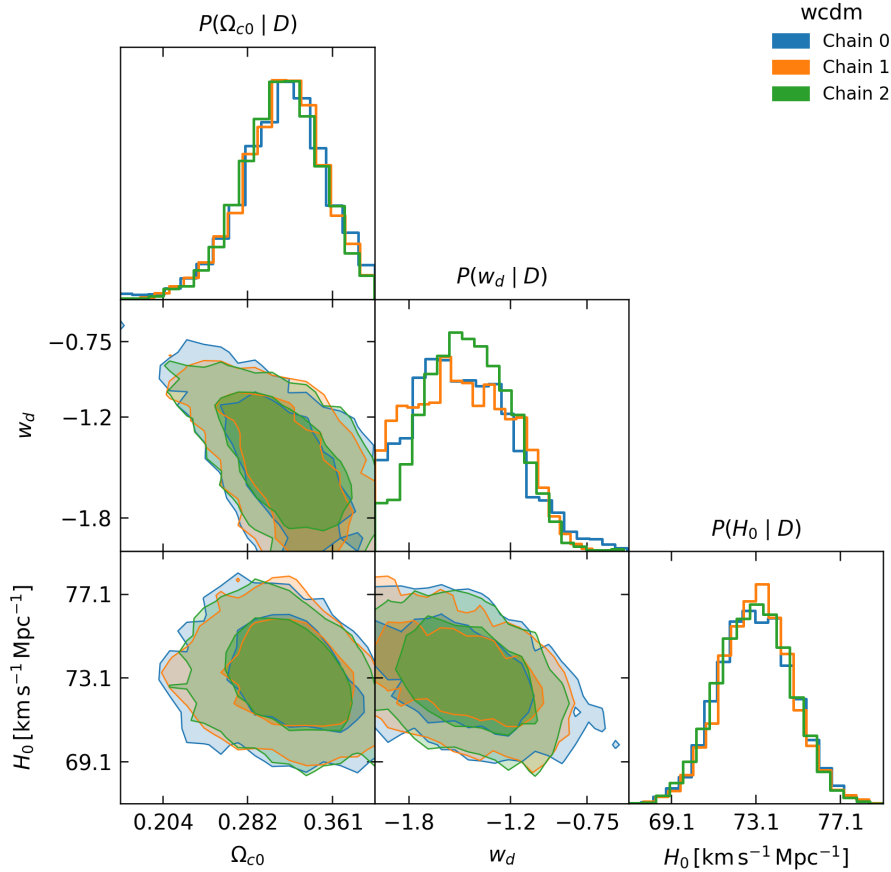

\centering
\capstart

\noindent\sphinxincludegraphics{{burst_plot}.png}
\caption{Triangle plot of separated chains from a simulation made with the command \sphinxcode{\sphinxupquote{burst}}.}\label{\detokenize{MCMC:id15}}\end{figure}

\subsection{MCMC from the GUI}
\label{\detokenize{GUIMCMC::doc}}\label{\detokenize{GUIMCMC:mcmc-from-the-gui}}
Starting with version 1.3, the simulations can also be performed from the
graphical interface.
Run:

\fvset{hllines={, ,}}%
\begin{sphinxVerbatim}[commandchars=\\\{\}]
\PYGZdl{} epic.py
\end{sphinxVerbatim}

to launch EPIC’s graphical interface.
When you choose and set up a model, instead of specifying the values of the
model parameters, you can now also infer them using MCMC directly from this
interface, where you can define the priors and choose the datasets to be used.

Select the tab “Constrain with MCMC” in the bottom frame.
Note that this area is made of three sections.
The left panel is where the priors are defined.
It is populated with the free parameters of the selected model after the model is
built.
The first column sets the priors, which can be flat or Gaussian, with two
parameters defining these distributions.
The entries are automatically filled with default values that the user should
change as needed.
Another option is to fix the parameter and indicate the fixed value.
By default all priors are set to flat, except the prior of the CMB temperature
when radiation is included, which is fixed but can also be changed.
The second column defines the square root of the variance of each parameter, to
compose a diagonal covariance matrix for the proposal function.
The third column shows the units of the quantities, when it is the case.

\begin{figure}[!tb]
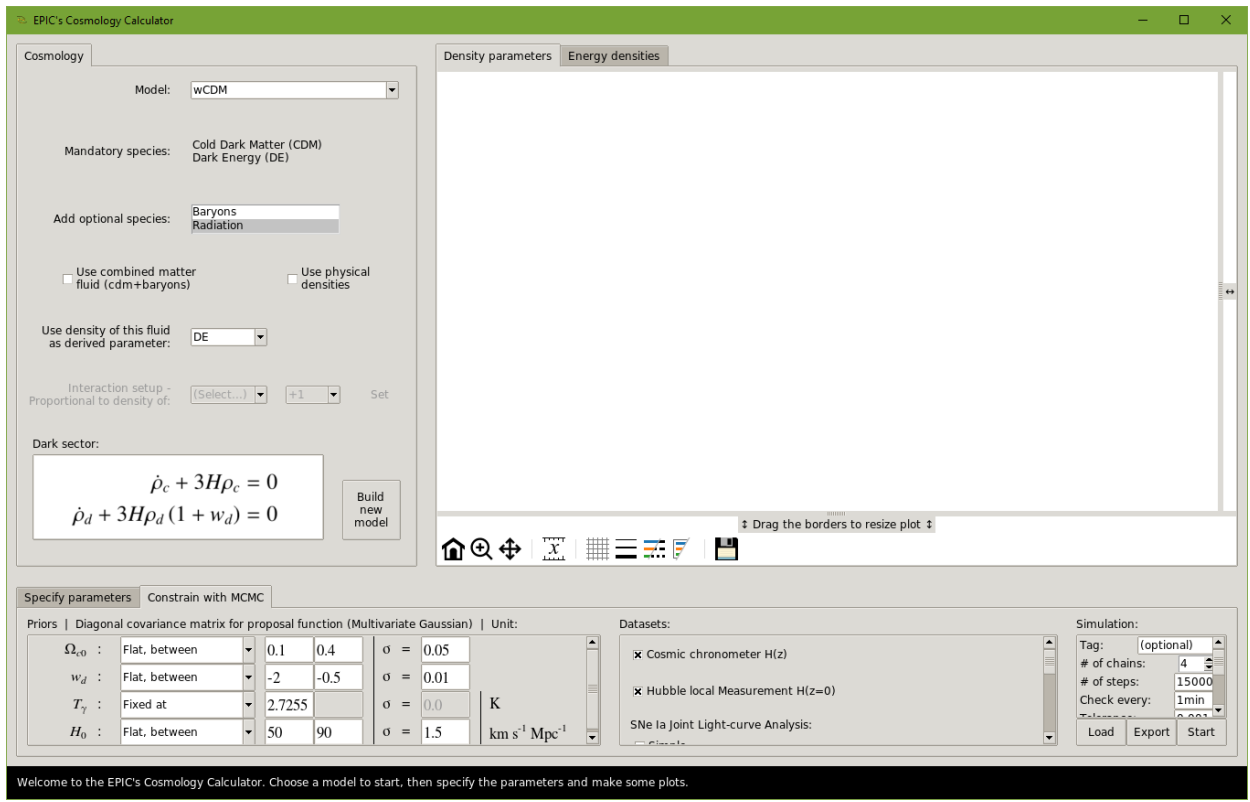

\centering
\capstart

\noindent\sphinxincludegraphics{{GUI-MCMC}.png}
\caption{MCMC with the Graphical User Interface.}\label{\detokenize{GUIMCMC:id1}}\end{figure}

The middle panel is where the user can choose which datasets to include in the
comparison (see section {\hyperref[\detokenize{thedata:datasec}]{\sphinxcrossref{\DUrole{std,std-ref}{The datasets}}}}).
Note that BAO and CMB data are only available when the model includes baryons
and radiation.
The JLA dataset (simple or full) will add more parameters to the analysis.
The priors for these nuisance parameters must as well be defined by the user.

Finally, the right panel defines options for the simulation, such as a tag for
the folder where the output will be saved, the number of chains, the number of
steps to be run in each loop, the time interval to check for convergence (and
display the plots), the tolerance for the convergence check, the desired
interval for the acceptance rates (which determines when chains should be
discarded), the number of sigma confidence levels to display in the plots and
report in the final output table, the number of bins for the histograms shown
during the evolution. Next, there are toggles to set whether or not, after
finishing the simulation, a last analysis with the \sphinxcode{\sphinxupquote{-{-}kde}} option should be
made, save the plots in \sphinxcode{\sphinxupquote{.png}} besides \sphinxcode{\sphinxupquote{.pdf}} (always saved) and to use a
proposal covariance matrix (which need not be diagonal) imported from a text
file.
There are also buttons enabling plot customization options much like those
available to the Cosmology Calculator.
All changes of style are applied to existing plots, which also affect any other
plots of the Cosmology Calculator (energy densities, density parameters or
distances) that the user may already have generated previously if the selected
option is available in the corresponding Calculator menus in the “Specify
Parameters” tab.

The “Export” button prepares a \sphinxcode{\sphinxupquote{.ini}} file based on these settings and can
also be used to run MCMC from the terminal (instructions are given in the
header of the \sphinxcode{\sphinxupquote{.ini}} file).
The “Start” button exports the \sphinxcode{\sphinxupquote{.ini}} file and starts the simulation right
away, after the user have informed the location to save the file.
When the simulation starts, a terminal output is shown in a new tab “Output” in
the bottom frame.
Also shown are the acceptance rates, the number of accepted states of each
chain (as the data are saved to the disk at each loop) and the normal fits to
the marginalized distributions (when convergence is checked).

\begin{figure}[!tb]
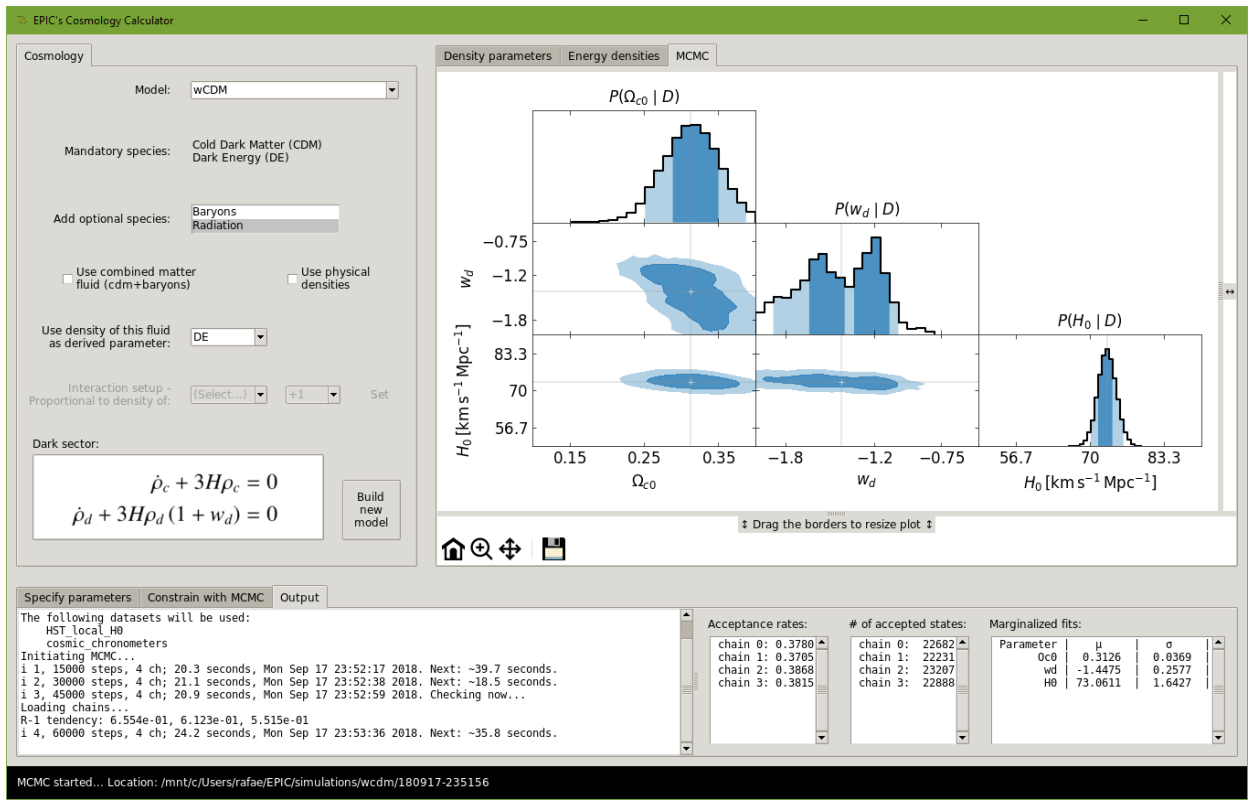

\centering
\capstart

\noindent\sphinxincludegraphics{{GUI-MCMC-running}.png}
\caption{MCMC with the Graphical User Interface, simulation running.}\label{\detokenize{GUIMCMC:id2}}\end{figure}

When finished, the plot is done again using ranges automatically adapted for
optimal view and can be further customized.

\begin{figure}[!tb]
\centering
\capstart

\noindent\sphinxincludegraphics{{GUI-MCMC-final}.png}
\caption{MCMC with the Graphical User Interface, final plot.}\label{\detokenize{GUIMCMC:id3}}\end{figure}

Starting with version 1.4, now you can also resume previously run simulations
in the GUI, regardless on whether this simulation was first run from the GUI or
from the terminal.
Just click the “Load” button, choose the folder corresponding to the simulation
that you want to resume and it will continue running right away.
Note that settings like the tolerance and interval for convergence checking
must be configured before clicking the “Load” button.

\section{Acknowledgments}
\label{\detokenize{acknowledgments::doc}}\label{\detokenize{acknowledgments:acknowledgments}}
I want to thank Elcio Abdalla, Gabriel Marcondes and Luiz Irber for their help.
This work has made use of the computing facilities of the Laboratory of
Astroinformatics (IAG/USP, NAT/Unicsul), whose purchase was made possible by
the Brazilian agency FAPESP (grant 2009/54006-4) and the INCT-A.

\renewcommand{\indexname}{Index}
\printindex
\end{document}